\documentclass{aa}

\usepackage{graphicx}
\usepackage{txfonts}
\usepackage{natbib}     
\usepackage{xcolor} 
\usepackage[para]{threeparttable} 
\usepackage{float}
\usepackage{threeparttable} 
\usepackage{placeins} 
\usepackage{multirow} 
\usepackage{tablefootnote}

\usepackage{caption}
\usepackage{subcaption}

\usepackage{natbib,twoopt} \usepackage[breaklinks=true]{hyperref}

\bibpunct{(}{)}{;}{a}{}{,} 

\begin{document}

\title{Exploring the ultra-hot Jupiter WASP-178b}
\subtitle{Constraints on atmospheric chemistry and dynamics from a joint retrieval of VLT/CRIRES$^+$ and space photometric data}

\author{
    D.~Cont \inst{\ref{instLMU}, \ref{instExzO}}
    \and L.~Nortmann \inst{\ref{instIAG}}
    \and F.~Yan \inst{\ref{instFei}}
    \and F.~Lesjak\inst{\ref{instIAG}}
    \and S.~Czesla \inst{\ref{instTLS}}
    \and A.~Lavail \inst{\ref{instIRAP}}    
    \and A.~Reiners\inst{\ref{instIAG}}
    \and N.~Piskunov \inst{\ref{instUppsala}}
    \and A.~Hatzes \inst{\ref{instTLS}}
    \and L.~Boldt-Christmas \inst{\ref{instUppsala}}
    \and O.~Kochukhov \inst{\ref{instUppsala}}
    \and T.~Marquart \inst{\ref{instUppsala}}   
    \and E.~Nagel\inst{\ref{instIAG}}   
    \and A.~D.~Rains \inst{\ref{instUppsala}}
    \and M.~Rengel \inst{\ref{instMPS}}
    \and U.~Seemann \inst{\ref{instESO}}
    \and D.~Shulyak \inst{\ref{instGrandada}}
    }      

\institute{
    Universit\"ats-Sternwarte, Ludwig-Maximilians-Universit\"at M\"unchen, Scheinerstrasse 1, 81679 M\"unchen, Germany\label{instLMU}\\ 
    \email{david.cont@lmu.de}
    \and
    Exzellenzcluster Origins, Boltzmannstrasse 2, 85748 Garching bei M\"unchen, Germany\label{instExzO}
    \and
    Institut f\"ur Astrophysik und Geophysik, Georg-August-Universit\"at G\"ottingen, Friedrich-Hund-Platz 1, 37077 G\"ottingen, Germany\label{instIAG} 
    \and
    Department of Astronomy, University of Science and Technology of China, Hefei 230026, People’s Republic of China\label{instFei}
    \and
    Th\"uringer Landessternwarte Tautenburg, Sternwarte 5, 07778 Tautenburg, Germany \label{instTLS}
    \and
    Institut de Recherche en Astrophysique et Plan\'etologie, Universit\'e de Toulouse, CNRS, IRAP/UMR 5277,
    14 avenue Edouard Belin, F-31400, Toulouse, France\label{instIRAP} 
    \and 
    Department of Physics and Astronomy, Uppsala University, Box 516, 75120 Uppsala, Sweden\label{instUppsala}   
    \and
    Max-Planck-Institut für Sonnensystemforschung, Justus-von-Liebig-Weg 3, 37077 G\"ottingen, Germany\label{instMPS}
    \and
    European Southern Observatory, Karl-Schwarzschild-Str. 2, 85748 Garching bei M\"unchen, Germany\label{instESO}
    \and 
    Instituto de Astrof{\'i}sica de Andaluc{\'i}a (IAA-CSIC), Glorieta de la Astronom{\'i}a s/n, 18008 Granada, Spain\label{instGrandada}
    }

\date{Received 21 March 2024 / Accepted 25 June 2024}


\abstract
{Despite recent progress in the spectroscopic characterization of individual exoplanets, the atmospheres of key ultra-hot Jupiters (UHJs) still lack comprehensive investigations. These include WASP-178b, one of the most irradiated UHJs known to date. We observed the dayside emission signal of this planet with CRIRES$^+$ in the spectral K band. By applying the cross-correlation technique and a Bayesian retrieval framework to the high-resolution spectra, we identified the emission signature of $^{12}$CO (S/N\,=\,8.9) and H$_2$O (S/N\,=\,4.9), and a strong atmospheric thermal inversion. A joint retrieval with space-based secondary eclipse measurements from TESS and CHEOPS allowed us to refine our results on the thermal profile and thus to constrain the atmospheric chemistry, yielding a solar to super-solar metallicity ($1.4\pm1.6$\,dex) and a solar C/O ratio ($0.6\pm0.2$). We infer a significant excess of spectral line broadening and identify a slight Doppler-shift between the $^{12}$CO and H$_2$O signals. These findings provide strong evidence for a super-rotating atmospheric flow pattern and suggest the possible existence of chemical inhomogeneities across the planetary dayside hemisphere. In addition, the inclusion of photometric data in our retrieval allows us to account for stellar light reflected by the planetary atmosphere, resulting in an upper limit on the geometric albedo (0.23). The successful characterization of WASP-178b's atmosphere through a joint analysis of CRIRES$^+$, TESS, and CHEOPS observations highlights the potential of combined studies with space- and ground-based instruments and represents a promising avenue for advancing our understanding of exoplanet atmospheres.}


\maketitle

%

\section{Introduction}

Ultra-hot Jupiters (UHJs) are a class of gas giant exoplanets with dayside temperatures exceeding 2200\,K \citep{Parmentier2018}. The separations between these planets and their host stars are within a few percent of the Earth-Sun distance, resulting in short orbital periods of a few hours to days. UHJs are typically located on orbits with negligible eccentricities and are thought to exhibit synchronous rotation, both of which are consequences of tidal circularization during the early evolution of these planetary systems \citep{Hut1981}. Since the planetary rotation period matches the orbital period, UHJs have permanent day- and nightsides with significant differences in thermal properties. Under the extreme thermal conditions of the dayside atmosphere, most molecular species are predicted to dissociate, limiting complex molecular chemistry to the nightside hemisphere \citep[e.g.,][]{Kitzmann2018, Helling2019}. In addition, theoretical work suggests that the presence of different chemical regimes between the two planetary hemispheres plays an important role in the global heat redistribution of UHJs \citep{BellCowan2018, KomacekTan2018}.

\begin{sidewaystable*}
\caption{Summary of chemical species detected via high-resolution spectroscopy in UHJ atmospheres, retrieved from the {\tt Exoplanet Atmospheres Database} as of May 2024.}             
\label{tab:summary-chemical-species}                           
\centering                                       
\renewcommand{\arraystretch}{2} 
\begin{threeparttable}
\small 
\begin{tabular}{l c c c c c c}                       
\hline\hline                             
\noalign{\smallskip}
Planet & $T_\mathrm{eq}(\mathrm{planet})$ & $T_\mathrm{eff}(\mathrm{star})$ & Neutral atoms & Ions & Molecules & References \\     
\noalign{\smallskip}
\hline                                   
\noalign{\smallskip}
HAT-P-70b            & 2562\,K & 8450\,K    & \parbox{5.3cm}{\centering Ca, Cr, Fe, H, Mg, Mn, Na, V} & \parbox{5cm}{\centering Ca$^+$, Cr$^+$, Fe$^+$, Ti$^+$} & & 1, 2 \\
KELT-9b              & 3921\,K & 9329\,K    & \parbox{5.4cm}{\centering Ca, Co, Cr, Fe, H, Mg, Na, Ni, O, Si, Ti, V} & \parbox{5cm} {\centering Ba$^+$, Ca$^+$, Cr$^+$, Fe$^+$, Na$^+$, Sc$^+$, Sr$^+$, Tb$^+$, Ti$^+$, Y$^+$ }  & & 3--18 \\ 
MASCARA-1b           & 2594\,K & 7554\,K    & \parbox{5.3cm}{\centering Cr, Fe, Ti} & & \parbox{4.6cm}{\centering CO, H$_2$O} & 19--22 \\ 
 MASCARA-2b           & 2261\,K & 8652\,K   & \parbox{5.3cm}{\centering Ca, Cr, Fe, H, Mg, Mn, Na, Ni, Si}  & \parbox{5cm}{\centering Ca$^+$, Cr$^+$, Fe$^+$} & & 2, 11, 12, 23--34 \\ 
MASCARA-4b           & 2250\,K & 7810\,K    & \parbox{5.3cm}{\centering Ca, Cr, Fe, H, Mg, Mn, Na, Ni, Rb, Sm, V} & \parbox{5cm}{\centering Ba$^+$, Ca$^+$, Fe$^+$, Sc$^+$, Ti$^+$} & & 2, 35, 36 \\ 
TOI-1518b            & 2492\,K & 7300\,K    & \parbox{5.3cm}{\centering Fe} & & & 37 \\ 
WASP-12b             & 2580\,K & 6313\,K    & \parbox{5.3cm}{\centering H, Na} & & & 38 \\ 
WASP-18b             & 2411\,K & 6400\,K    & & & \parbox{4.6cm}{\centering CO, H$_2$O, OH} & 39, 40 \\ 
WASP-19b             & 2077\,K & 5460\,K    & & & \parbox{4.7cm}{\centering TiO} & 41 \\ 
WASP-33b             & 2710\,K & 7430\,K    & \parbox{5.3cm}{\centering Fe, H, Si, Ti, V} & \parbox{5cm}{\centering Ca$^+$, Ti$^+$} & \parbox{4.6cm}{ \centering CO, H$_2$O, OH, TiO} & 6, 30, 42--54 \\ 
WASP-76b             & 2228\,K & 6329\,K    & \parbox{5.3cm}{\centering Ca, Co, Cr, Fe, H, K, Li, Mg, Mn, Na, Ni, O, V} & \parbox{5cm}{\centering Ba$^+$, Ca$^+$, Fe$^+$, Sr$^+$} & \parbox{4.6cm}{\centering CO, HCN, H$_2$O, OH ,VO} & 2, 11, 39, 55--72 \\ 
WASP-121b            & 2358\,K & 6586\,K    & \parbox{5.2cm}{\centering Ca, Co, Cr, Fe, H, K, Li, Mg, Mn, Na, Ni, V} & \parbox{4.6cm}{\centering Ba$^+$, Ca$^+$, Fe$^+$, Sc$^+$, Sr$^+$} & & 2, 11, 63, 73--82 \\ 
WASP-178b            & 2470\,K & 8640\,K    & \parbox{5.3cm}{\centering H, Fe, Mg, Na} & \parbox{5cm}{\centering Fe$^+$} & \parbox{4.6cm}{\centering CO, H$_2$O} & this work, 83 \\ 
WASP-189b            & 2641\,K & 8000\,K    & \parbox{5.3cm}{\centering Ca, Cr, Fe, H, Mg, Mn, Na, Ni, Sr, Ti, V} & \parbox{5cm}{\centering Ba$^+$, Ca$^+$, Fe$^+$, Sr$^+$, Ti$^+$} &  \parbox{4.6cm}{\centering CO, TiO} & 2, 11, 50, 84--88 \\                                                   
\noalign{\smallskip}
\hline   
\end{tabular}
\vspace{0.3cm}
\begin{tablenotes}
\textbf{References. }  
(1)~\cite{Bello-Arufe2022a}, (2)~\cite{Gandhi2023}, 
(3)~\cite{Hoeijmakers2018}, (4)~\cite{Hoeijmakers2019}, (5)~\cite{Yan2018}, (6)~\cite{Yan2019}, (7)~\cite{Wyttenbach2020}, (8)~\cite{Kasper2021}, (9)~\cite{Pino2022}, (10)~\cite{Sanchez-Lopez2022b}, (11)~\cite{Langeveld2022}, (12)~\cite{Bello-Arufe2022b}, (13)~\cite{Pai-Asnodkar2022}, (14)~\cite{Borsa2022b}, (15)~\cite{Borsato2023a}, (16)~\cite{Borsato2023b}, (17)~\cite{Ridden-Harper2023}, (18)~\cite{Lowson2023}, 
(19)~\cite{HolmbergMadhusudhan2022}, (20)~\cite{Ramkumar2023}, (21)~\cite{Scandariato2023}, (22)~\cite{Guo2024}, 
(23)~\cite{Casasayas-Barris2018}, (24)~\cite{Casasayas-Barris2019}, (25)~\cite{Hoeijmakers2020a}, (26)~\cite{Nugroho2020a}, (27)~\cite{Stangret2020}, (28)~\cite{Borsa2022a}, (29)~\cite{Yan2022a}, (30)~\cite{Cont2022a}, (31)~\cite{Fossati2023}, (32)~\cite{Kasper2023}, (33)~\cite{Johnson2023}, (34)~\cite{Petz2024}, 
(35)~\cite{Zhang2022}, (36)~\cite{Jiang2023}, 
(37)~\cite{Cabot2021}, 
(38)~\cite{Jensen2018}, 
(39)~\cite{Yan2023}, (40)~\cite{Brogi2023} 
(41)~\cite{Sedaghati2021}, 
(42)~\cite{Nugroho2017}, (43)~\cite{Nugroho2020b}, (44)~\cite{Borsa2021a}, (45)~\cite{Yan2021}, (46)~\cite{Cont2021}, (47)~\cite{Nugroho2021}, (48)~\cite{Cont2022b}, (49)~\cite{Herman2022}, (50)~\cite{Yan2022b}, (51)~\cite{Finnerty2023}, (52)~\cite{vanSluijs2023}, (53)~\cite{Wright2023}, (54)~\cite{Yang2024},  
(55)~\cite{Seidel2019}, (56)~\cite{Ehrenreich2020}, (57)~\cite{Landman2021}, (58)~\cite{Deibert2021}, (59)~\cite{Casasayas-Barris2021}, (60)~\cite{Seidel2021}, (61)~\cite{Kesseli2021}, (62)~\cite{Tabernero2021}, (63)~\cite{Azevedo-Silva2022}, (64)~\cite{Kesseli2022}, (65)~\cite{Sanchez-Lopez2022a}, (66)~\cite{Gandhi2022}, (67)~\cite{Kawauchi2022}, (68)~\cite{Deibert2023}, (69)~\cite{Pelletier2023}, (70)~\cite{Gandhi2024}, (71)~\cite{Maguire2024}, (72)~\cite{WeinerMansfield2024}, 
(73)~\cite{Hoeijmakers2020b}, (74)~\cite{Ben-Yami2020}, (75)~\cite{Cabot2020}, (76)~\cite{Merritt2021}, (77)~\cite{Borsa2021b}, (78)~\cite{Hoeijmakers2022}, (79)~\cite{Gibson2022}, (80)~\cite{Seidel2023}, (81)~\cite{Maguire2023}, (82)~\cite{Young2024}, 
(83)~\cite{Damasceno2024}, 
(84)~\cite{Yan2020}, (85)~\cite{Stangret2022}, (86)~\cite{Prinoth2022}, (87)~\cite{Prinoth2023}, (88)~\cite{Prinoth2024}. 
\end{tablenotes}            
\end{threeparttable}
\end{sidewaystable*}

In recent years, observational work has devoted considerable effort to exploring the properties of UHJ atmospheres. In particular, spectroscopic studies of the atmospheric composition have revealed the presence of features from a wide variety of chemical species. Table~\ref{tab:summary-chemical-species} gives an overview of the chemical species detected with high-resolution spectroscopy in UHJ atmospheres, along with the names of the corresponding planets\footnote{Retrieved from the {\tt Exoplanet Atmospheres Database} at \url{https://research.iac.es/proyecto/exoatmospheres/}}. So far, the chemical inventory of UHJ atmospheres has been extensively studied by both transmission and emission spectroscopy at visible wavelengths, which are dominated by spectral lines originating from atomic and ionic metal species \citep[e.g.,][]{Hoeijmakers2019, Tabernero2021, Kesseli2022, Pino2020, Kasper2021}. In this context, emission spectroscopy has proven to be a powerful tool for the identification of inverted temperature-pressure ($T$-$p$) profiles in the dayside atmospheres of UHJs \citep[e.g.,][]{Nugroho2017, Cont2021, Kasper2021, Finnerty2023}. These so-called thermal inversions correspond to a temperature pattern that increases with altitude and are caused by strong absorption of incoming stellar radiation at visible and ultraviolet wavelengths in the upper planetary atmosphere. Initially, this absorption mechanism was attributed exclusively to the presence of TiO and VO \citep{Hubeny2003, Fortney2008}, but subsequent theoretical and observational work has shown that neutral and ionized atomic metals may also be fundamental for maintaining the thermal inversion layers \citep{Lothringer2018, Arcangeli2018, Gandhi2019_2, Piette2020}.


In addition, recent observations have revealed prominent emission features of molecular species in the infrared spectral range, particularly of CO, H$_2$O, and the hydroxyl radical (OH; \citealt{Nugroho2021, Yan2022b, HolmbergMadhusudhan2022, Finnerty2023, Yan2023, Brogi2023}). Species with strong chemical bonds (e.g., CO) dominate the inventory of molecular species in the hottest atmospheric regions due to their ability to withstand extreme temperatures and elevated irradiation levels. In contrast, the cooler regions of UHJ atmospheres are also populated by molecules with more moderate binding energies (e.g., H$_2$O; \citealt{Parmentier2018, Kreidberg2018}). The molecular species CO, H$_2$O, and OH harbor a significant fraction of atmospheric carbon and oxygen, enabling the estimation of the carbon-to-oxygen (C/O) abundance ratio. This quantity has been proposed as an important tracer of planet formation and migration history, making observations at infrared wavelengths a critical task for studying planetary evolution pathways \citep[e.g.,][]{Oeberg2011, Madhusudhan2012, Mordasini2016}.

High-resolution spectroscopy observations in the infrared have allowed precise studies of atmospheric CO and H$_2$O, the dominant carbon- and oxygen-bearing species of UHJ atmospheres (e.g., WASP-18b, WASP-76b, MASCARA-1b; \citealt{Yan2023, Brogi2023, Ramkumar2023, Hood2024}). These studies are based on the cross-correlation technique, which relies on the Doppler-shift of a planet's orbital motion relative to the telluric and stellar lines to identify its spectral signature \citep[e.g.,][]{Snellen2010, Brogi2012, Birkby2018}. This method has proven to be a powerful tool for detecting the spectral signature of exoplanet atmospheres \citep[e.g.,][]{Sanchez-Lopez2019, Yan2020, Cont2022b}. However, the cross-correlation method lacks the ability to extract quantitative constraints on atmospheric parameters in a straightforward way.

To overcome this difficulty, atmospheric retrieval frameworks have been developed to fit high-resolution spectroscopy observations with parameterized model spectra \citep{Brogi-Line2019, Gibson2020}. Mostly, these frameworks use Bayesian methods to estimate the best-fit values and uncertainties of the model parameters given the observed spectrum of an exoplanet's atmosphere. Recent advances in Bayesian inference techniques, combined with high-quality observational data, have enabled detailed investigations of fundamental atmospheric properties, such as $T$-$p$ profiles or elemental abundances. In particular, recent studies of the hot Jupiter \mbox{WASP-77Ab} highlight the great potential of retrieval techniques, allowing the atmospheric metallicity and C/O ratio to be constrained with unprecedented precision, providing crucial insights into the planet's migration history \citep{Line2021, Smith2024}. Many efforts have been made in exoplanet retrieval codes \citep{RengelAdamczewski2023, MacDonald2023}, and further development of high-resolution spectroscopy retrieval frameworks is currently underway. These improvements include the use of more comprehensive $T$-$p$ profiles \citep[e.g.,][]{Kitzmann2020, Pelletier2021}, or pressure-dependent mixing ratios of chemical species instead of the predominantly assumed constant vertical abundances \citep[e.g.,][]{Cont2022b, Pelletier2023}.

\begin{table}
        \caption{Parameters of the \object{WASP-178} system.}             
        \label{tab-parameters}                           
        \centering                                       
        \renewcommand{\arraystretch}{1.3} 
        \begin{threeparttable}
                \begin{tabular}{l c c}                       
                \hline\hline                             
                \noalign{\smallskip}
                Parameter & Symbol (Unit) & Value \\     
                \noalign{\smallskip}
                \hline                                   
                \noalign{\smallskip}
                \textit{Planet} & &  \\ 
                \noalign{\smallskip}
                Radius                              & $R_\mathrm{p}$ ($R_\mathrm{Jup}$) & $1.940_{-0.058}^{+0.060}$ \\
                Orbital period                      & $P_\mathrm{orb}$ (d) & $3.3448412$ \\
                Transit epoch  & $T_\mathrm{0}$ (BJD$_\mathrm{TDB}$) &  $2458321.86724$ \\ 
                Transit duration     & $T_\mathrm{transit}$ (h) & $3.63 \pm 0.04$ \\                        
                Ingress duration     & $T_\mathrm{ingress}$ (h) & $0.57_{-0.03}^{+0.04}$ \\
                Orbital inclination     & $i$ (deg) & $84.45$ \\                         
                Surface gravity  & $\log{g}$ (cgs) & $2.97$ \\
                RV semi-amplitude    & $K_\mathrm{p}$ (km\,s$^{-1}$)       & $176.5$ \\
                                                     &                                     & $181.0$ $^{(*)}$ \\                       
                Systemic velocity                   & $\varv_\mathrm{sys}$ (km\,s$^{-1}$) & $-25.595_{-0.024}^{+0.025}$ \\
                                                     &                                     & $-23.908\pm0.007$ $^{(*)}$ \\
                \noalign{\smallskip} \hline \noalign{\smallskip}
                \textit{Star} & &  \\  
                \noalign{\smallskip}
                Radius & $R_*$ ($R_\mathrm{\sun}$) & $1.801_{-0.048}^{+0.049}$ \\ 
                Effective temperature & $T_\mathrm{eff}$ (K) & $8640_{-240}^{+500}$ \\          
                \noalign{\smallskip}
                \hline                    
                \end{tabular}
                \vspace{0.3cm}
                \begin{tablenotes}
                \textbf{Notes.} $^{(*)}$ \cite{Hellier2019}, all other values from \cite{Rodriguez-Martinez2020}. The RV semi-amplitude values were calculated using the expression, 
                \mbox{$K_\mathrm{p} = \left( 2 \pi G  M_* / P_\mathrm{orb} \right)^{1/3} \cdot \sin{i}$} 
                where $G$ is the gravitational constant and all other parameters are listed in the table. In the present study we use the values from \cite{Rodriguez-Martinez2020} due to the more recent transit epoch. The values of $K_\mathrm{p}$ and $\varv_\mathrm{sys}$ obtained from \cite{Hellier2019} are shown here for comparison.
                \end{tablenotes} 
        \end{threeparttable}
\end{table}

In addition to deriving information about atmospheric thermal structures and chemical abundances, cross-correlation and retrieval frameworks offer the possibility to study global circulation and constrain planetary rotation parameters. For example, recent observations with IGRINS and CRIRES$^+$ of the infrared CO, H$_2$O, OH and \ion{Fe}{i} signatures in the spectra of WASP-18b and MASCARA-1b have revealed Doppler-shifts in disagreement between the different chemical species \citep{Brogi2023, Ramkumar2023}. A similar effect was observed at visible wavelengths with CARMENES, which found evidence for a significant Doppler-shift between the emission signatures of \ion{Fe}{i} and TiO in the atmosphere of WASP-33b \citep{Cont2021}. These offsets are most likely the result of inhomogeneous distributions of the chemical species, combined with the effects of planetary rotation and winds. Atmospheric layers at different altitudes, probed by the spectral lines of different chemical species, could also contribute to the observed Doppler-shifts. Furthermore, studying the spectral line shape can reveal details about planetary rotation and atmospheric super-rotation. This concept, originally proposed by \cite{Snellen2010}, is supported by successful retrievals of the rotational broadening profile in a number of exoplanet spectra (e.g., WASP-33b, WASP-18b, WASP-76b, WASP-43b, $\beta$~Pictoris~b, HIP~65Ab; \citealt{Cont2022b, Yan2023, Lesjak2023, Landman2023, Bazinet2024}).

This study provides the first spectroscopic characterization of the dayside emission spectrum of the UHJ WASP-178b. We report detections of the spectral signatures of the molecular species $^{12}$CO and H$_2$O. In addition, we present the parameters of the planetary atmosphere obtained using a Bayesian retrieval framework. The discovery of WASP-178b, orbiting an A1-type star with a period of $P_\mathrm{orb}$\,$\sim$\,3.3\,d and a planetary equilibrium temperature of $T_\mathrm{eq}$\,$\sim$\,2500\,K, was announced independently by \cite{Hellier2019} and \cite{Rodriguez-Martinez2020}. The current literature on the planet is limited to a small number of transmission spectroscopy and photometric observations. \cite{Lothringer2022} detected significant absorption in the ultraviolet wavelength range at low spectral resolution with HST/WFC3, possibly caused by Si- and Mg-bearing cloud precursor species. Very recent high-resolution ESPRESSO observations also revealed the presence of the \ion{Fe}{i}, \ion{Fe}{ii}, \ion{Mg}{i} and \ion{Na}{i} spectral signatures, and the H$\alpha$ line at the planetary terminator \citep{Damasceno2024}. Finally, \cite{Pagano2023} used TESS and CHEOPS to determine the transit and eclipse depths, and to derive constraints on the reflection properties of the planet. All parameters of the WASP-178 system used in this work are summarized in the Table~\ref{tab-parameters}.

We structure this work as follows. The observations and data reduction procedures are described in Sects.~\ref{Observations} and \ref{Data reduction}. The technique for identifying the spectral lines of different chemical species and the resulting detections are outlined in Sect.~\ref{Detection of the planetary emission lines}. Our retrieval framework and the derived parameters of WASP-178b's atmosphere are described and discussed in Sect.~\ref{Retrieval of the atmospheric properties}. Finally, we give conclusions about our work in Sect.~\ref{Conclusions}.

%

\begin{figure}
\centering
\includegraphics[width=\columnwidth]{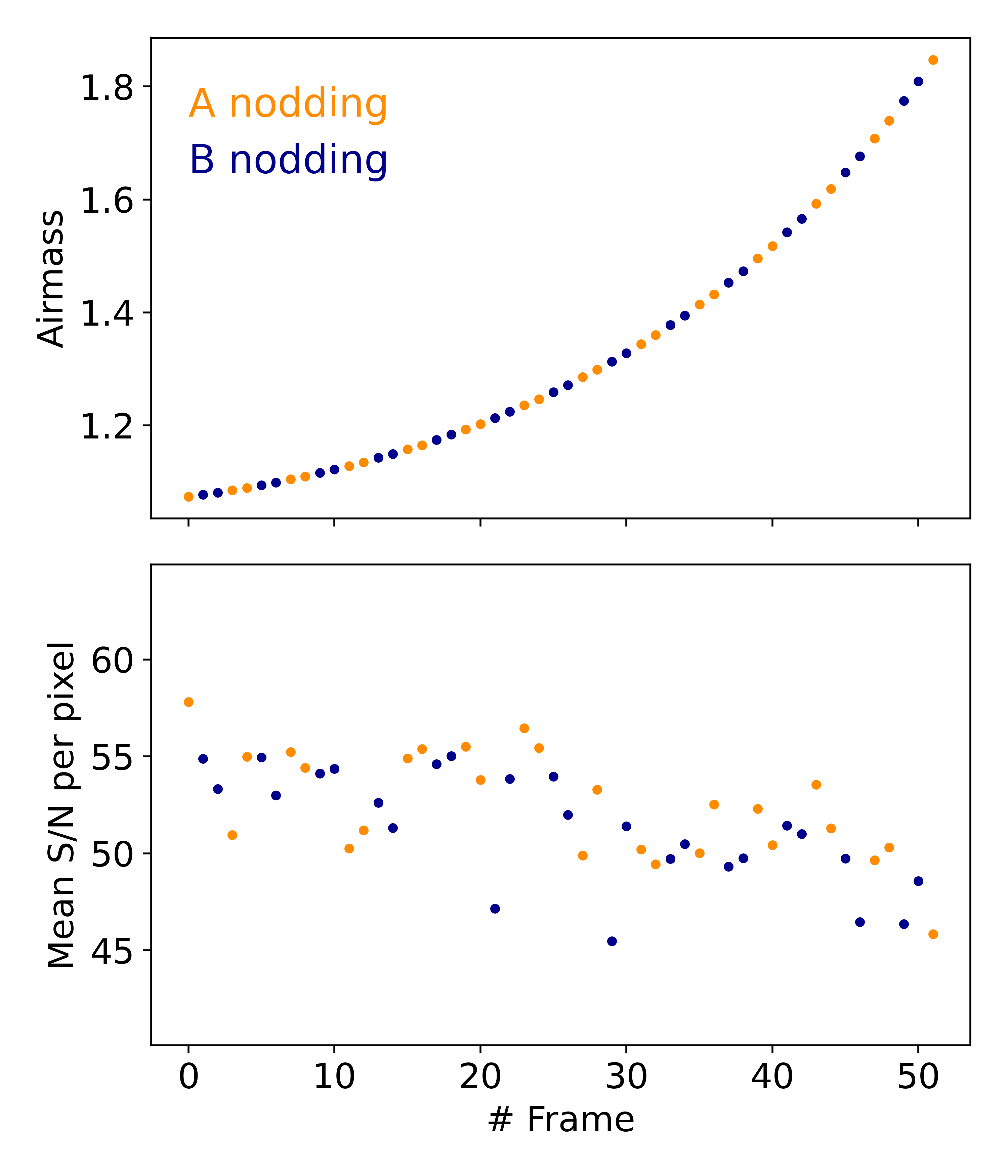}
\caption{Evolution of the airmass ({\it top panel}) and the mean S/N per detector pixel ({\it bottom panel}). Data points corresponding to the A and B nodding positions are shown in orange and blue, respectively.}
\label{figure:airmass-SNR}
\end{figure}

\section{Observations}
\label{Observations}

We observed the thermal emission spectrum of WASP-178b on 25 May 2023 between 04:35\,UT and 08:13\,UT with CRIRES$^+$ \citep{Dorn2023}. The CRIRES$^+$ instrument is a cryogenic cross-dispersed high-resolution near-infrared spectrograph, installed at the 8-meter Unit Telescope 3 of the Very Large Telescope. The observations of the planet cover the orbital phase interval 0.52--0.57, when the dayside of WASP-178b is almost completely aligned with the observer's line of sight. A total of 52 consecutive spectra were acquired with an integration time of 240\,s. We used an ABBA nodding pattern, which consists of observing the target at two different slit positions (A and B), to facilitate the removal of the sky background and detector artifacts in the subsequent data reduction steps. The slit width was set to 0.2$^{\prime\prime}$, giving a nominal resolving power of $R$\,$\sim$\,100\,000 under the condition of a homogeneous slit illumination. In our observations, however, the performance of the adaptive optics system was very good, resulting in a point spread function (PSF) smaller than the slit, and thus an increased spectral resolution. From the width of the PSF, we determined that the spectral resolution of our observations was $R$\,$\sim$\,135\,000. Our spectra were acquired in the K2166 wavelength setting, which offers a wavelength coverage between 1921~nm and 2472~nm. This setting consists of seven echelle orders, each divided by two narrow gaps between detectors, yielding a total of 21 wavelength segments. The airmass varied between 1.1 and 1.8 from the start to the end of the observations and the mean signal-to-noise ratio (S/N) per detector pixel was 52 (Fig.~\ref{figure:airmass-SNR}).

%

\section{Data reduction}
\label{Data reduction}

\subsection{Extraction of the one-dimensional spectra}
\label{Extraction of the one-dimensional spectra}

We used the CRIRES$^+$ data reduction software pipeline {\tt cr2res} to extract the one-dimensional spectra from the raw frames\footnote{\url{https://www.eso.org/sci/software/pipelines/index.html\#pipelines_table}}. As a first step, we processed the raw calibration frames taken by the observatory as part of the regular daily calibration routine. These data consist of dark frames, flat fields, and frames taken with a Uranium-Neon lamp and a Fabry-Perot etalon. The calibration data were downloaded from the ESO data archive and reduced using the standard reduction cascade as described in the {\tt cr2res} pipeline user manual.

We then reduced the science spectra. We grouped the 52 spectra of our time series into 26 pairs of spectra taken at A and B nodding positions. Each nodding pair was reduced using the \verb!cr2res_obs_nodding! recipe. This procedure applies flat field normalization and bad pixel masking from the calibrations to the raw frames, removes the sky background and instrumental artifacts via subtraction of the A and B frames from each other, and optimally extracts the one-dimensional science spectra and their uncertainties for each wavelength segment.

In general, the wavelength solutions for the A and B spectra calculated by the instrument pipeline are based on observations of a calibration lamp that uniformly illuminates the slit. The wavelength solutions are therefore only accurate if the stellar PSF either fills the slit or is exactly aligned with its center. In our observations, the PSF was smaller than the slit and its position along the dispersion direction was not perfectly centered within the slit, causing an unknown shift of the spectral data points relative to the wavelength solutions of the pipeline. This shift generally differs for the A and B spectra due to the telescope's nodding motion not being perfectly parallel to the slit orientation of the spectrograph. Therefore, each nodding position had its own wavelength solution. 

To account for the shifts in the spectral data points relative to the wavelength solutions of the pipeline, an additional refinement step was applied. We calculated the mean spectrum at each nodding position and applied {\tt molecfit} \citep{Smette2015} to the resulting A and B master spectra. By fitting the telluric absorption lines in the master spectra, {\tt molecfit} corrected the input wavelength solutions by the respective shift and provided us with an improved solution for the A and B spectra as output\footnote{The extracted data, together with the refined wavelength solution, are available at  \url{https://zenodo.org/records/11637332} \citep{Lavail2024}}. Although the applied procedure results in different {\tt molecfit} wavelength solutions for A and B, we refrained from shifting the spectral data points to a common wavelength solution to avoid introducing spurious signals during the required interpolation process. Instead, following \cite{Yan2023}, we treated the A and B spectra as two independent data sets throughout the subsequent procedures of this section, and combined their information without the need for interpolation in Sects.~\ref{Detection of the planetary emission lines} and \ref{Retrieval of the atmospheric properties} after data reduction was complete.

\begin{figure}
        \centering
        \includegraphics[width=\columnwidth]{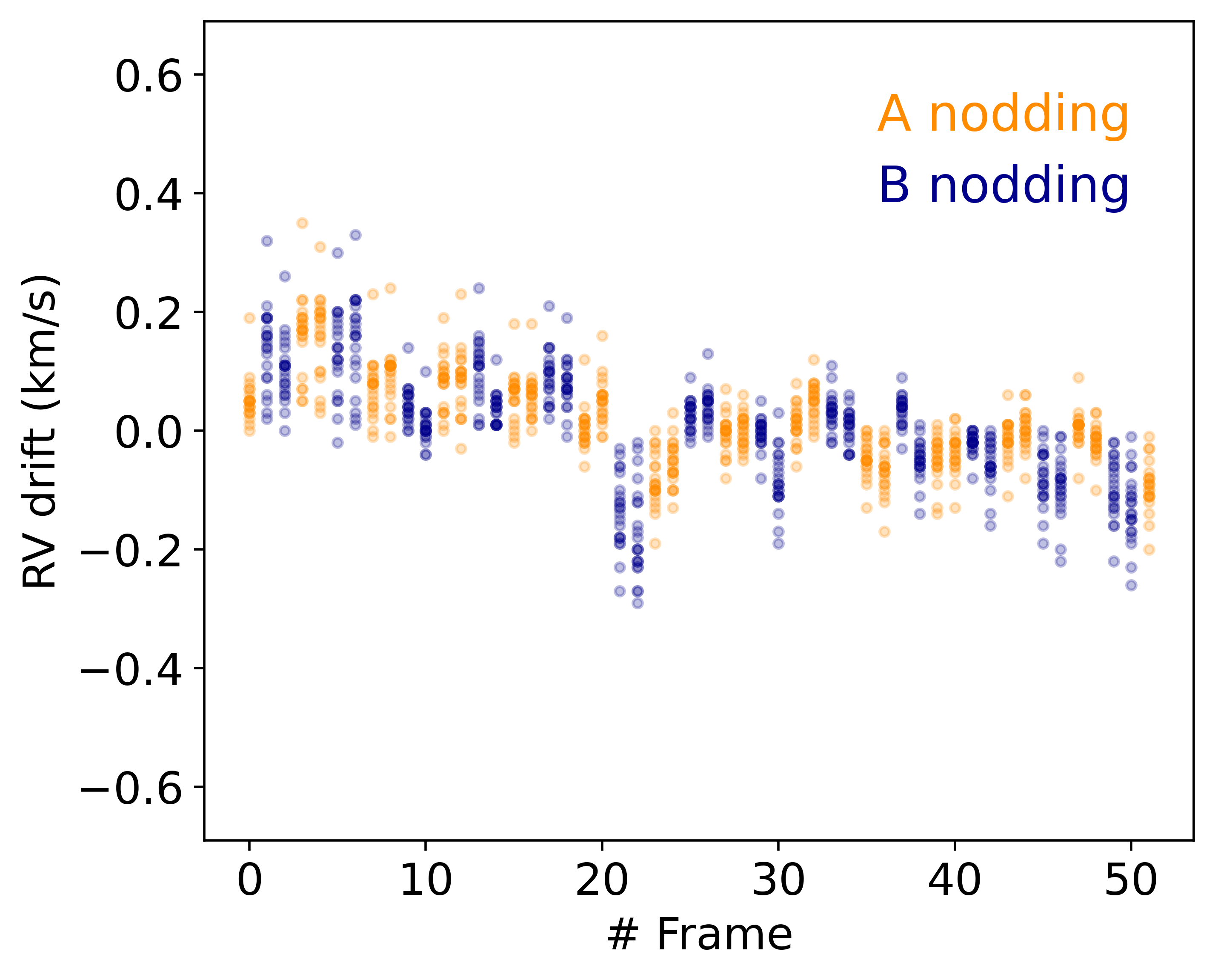}
        \caption{RV drift during the acquisition of the spectral time series. Each data point corresponds to an individual exposure frame and wavelength segment. Data points corresponding to the A and B nodding positions are shown in orange and blue, respectively.}
        \label{figure:RV-drift}
\end{figure}

\begin{figure}
        \centering
        \includegraphics[width=\columnwidth]{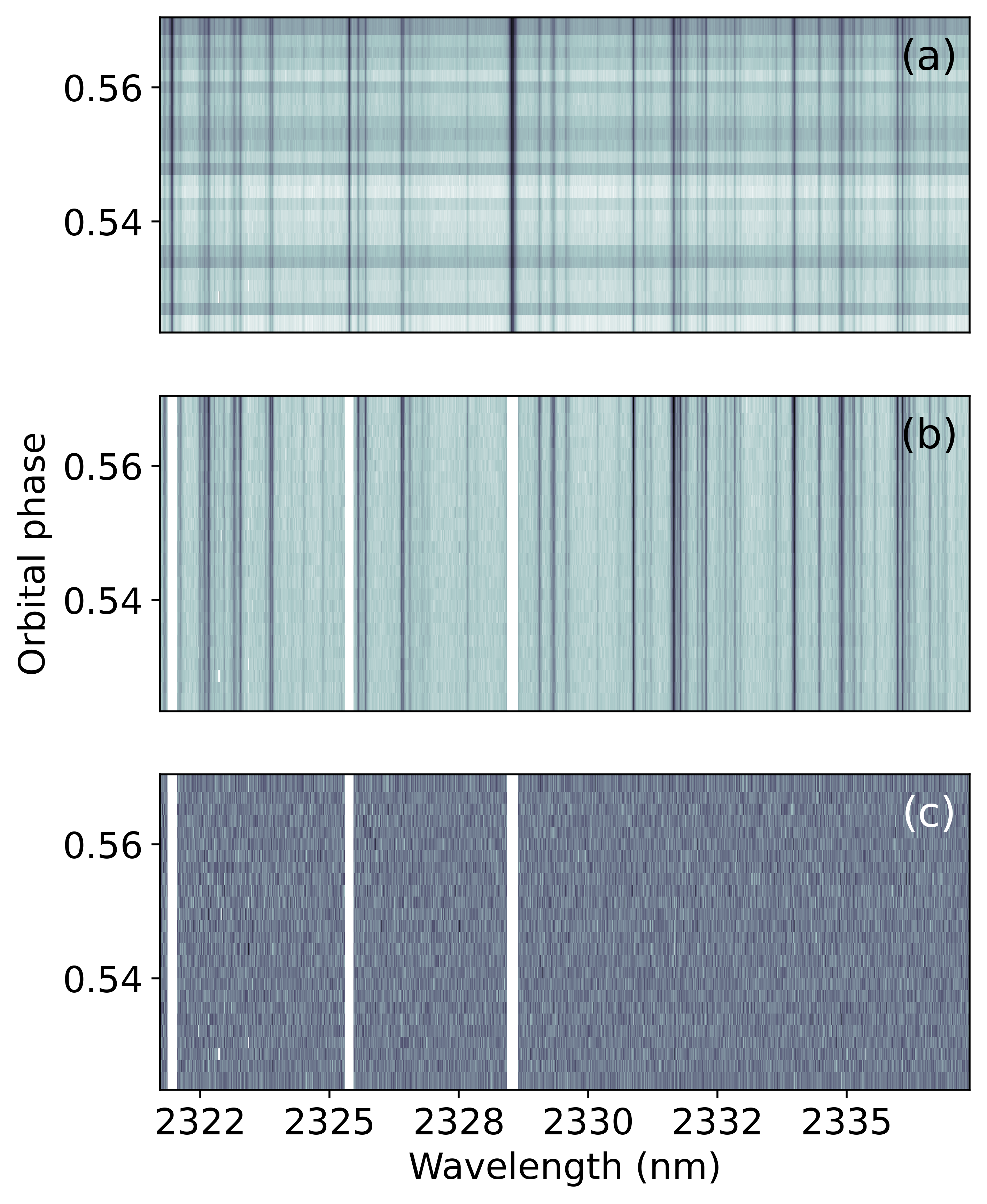}
        \caption{Overview of data reduction procedures for a representative CRIRES$^+$ wavelength range. { \it Panel a}: raw spectral matrix after extraction of the one-dimensional spectra. { \it Panel b}: spectral matrix after normalization, outlier correction, and masking of the strongest telluric features. {\it Panel c}: residual spectral matrix after removal of telluric and stellar lines via two consecutive {\tt SYSREM} iterations.}
        \label{figure:preprocessing-sysrem}
\end{figure}

\begin{figure*}
        \sidecaption
        \includegraphics[width=\textwidth]{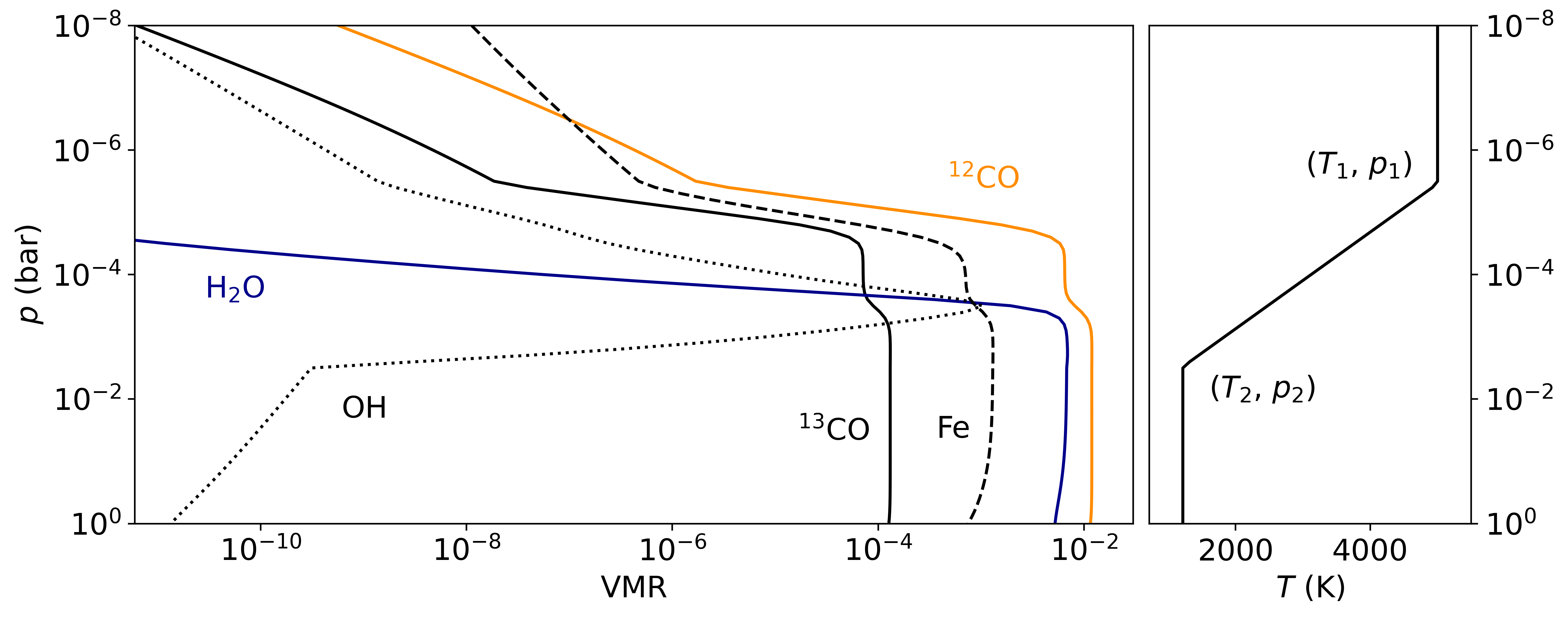}
        \caption{Volume mixing ratios (VMRs; {\it left panel}) and $T$-$p$ profile ({\it right panel}) used to generate the model spectra for cross-correlation. The VMRs were calculated assuming equilibrium chemistry. The elemental abundances and the $T$-$p$ profile used correspond to the best-fit results of the retrieval framework in Sect.~\ref{Retrieval of the atmospheric properties}.}
        \label{figure:VMR-Tp}
\end{figure*}




\subsection{Pre-processing the spectra}
\label{Pre-processing the spectra}

For each wavelength segment, the extracted one-dimensional spectra were sorted chronologically and stacked in a two-dimensional array, yielding a so-called spectral matrix. The baselines of the individual spectra differ from each other due to the variability of the atmospheric conditions during the observations. It is therefore necessary to normalize all spectra to the same continuum level: first, we calculated the median of all spectra in the time series and applied a second-order polynomial fit to the resulting master spectrum. In a second step, all spectra of the spectral matrix were divided by the master and then fitted with a first-order polynomial function. Finally, the continuum correction was obtained as the product of the two previously determined polynomial fit functions and applied by dividing each spectrum.

Outliers were removed using Principal Component Analysis. Using this method, we computed a model of the spectral matrix, subtracted the model from the data, and identified pixels that differed by 5$\sigma$ from the resulting residuals. The affected pixels were then removed from the spectral matrix. In addition, we masked all wavelength bins with a flux level below 30\% of the spectral continuum. We also masked wavelength bins with flux values above 120\% of the continuum level to account for strong sky emission lines.

We examined the temporal stability of our {\tt molecfit} wavelength solution from Sect.~\ref{Extraction of the one-dimensional spectra} by cross-correlating each exposure frame with the median spectrum of the time series. Figure~\ref{figure:RV-drift} shows the radial velocity (RV) drift of the spectra relative to the median for all wavelength segments. The RV drift between the start and end of the observations was less than 0.6\,km\,s$^{-1}$, with the majority of wavelength segments having even lower drifts on the order of 0.2\,km\,s$^{-1}$. Both offset values are significantly smaller than a single resolution element of the selected CRIRES$^+$ wavelength setting ($\sim$\,2.2\,km\,s$^{-1}$ at $R$\,=\,135\,000). Thus, the {\tt molecfit} wavelength solution was considered accurate enough for our purposes.

\subsection{Removal of telluric and stellar lines}
\label{Removal of telluric and stellar lines}

We used the detrending algorithm {\tt SYSREM} \citep{Tamuz2005} to correct for the contribution of telluric and stellar lines in the spectral data. Originally designed to identify and remove common signals in light curve studies, the algorithm has demonstrated to be a powerful tool in exoplanet research when applied to high-resolution spectroscopic time series \citep[e.g.,][]{Birkby2013, Nugroho2017, Sanchez-Lopez2019}. 

{\tt SYSREM} is usually applied over multiple iterations. The algorithm decomposes the contribution of systematic effects to the spectral time series into two column vectors $\mathbf{a}$ and $\mathbf{c}$, whose outer product results in a model of the systematics. This model is subtracted from the data, yielding a residual spectral matrix. Based on this residual spectral matrix, a new model with a new set of $\mathbf{a}$ and $\mathbf{c}$ is calculated. An updated residual spectral matrix is obtained by subtracting the new model from the result of the previous iteration step. The final model of systematic effects after $N$ iterations of {\tt SYSREM} is thus
\begin{equation}
    \mathbf{M}_\mathrm{systematic} = \sum_{k=1}^{N} \mathbf{a}_k {\mathbf{c}_k}^T = \mathbf{A} \mathbf{C}^T,
    \label{equation:sysrem}
\end{equation}
where $\mathbf{A}$ and $\mathbf{C}$ are matrices containing all column vectors $\mathbf{a}_k$ and $\mathbf{c}_k$. Subtracting the model obtained in Eq.~\ref{equation:sysrem} from the data yields the final residual spectral matrix, which is used in further analysis steps. We refer the reader to \cite{Czesla2024} for a more detailed mathematical description of {\tt SYSREM} in the context of exoplanet science.

The input provided to {\tt SYSREM} includes the pre-processed spectral matrix and the propagated uncertainties from the instrument pipeline. We applied the algorithm following the procedure described by \cite{Gibson2022}, which consists of first dividing the spectral matrix by the median spectrum and then subtracting the systematics modeled with {\tt SYSREM}. This approach allows the distortions of the planet's signal introduced by the algorithm to be accounted for when a Bayesian retrieval framework is applied to the data in Sect.~\ref{Retrieval of the atmospheric properties}, while preserving the relative strengths of the planetary spectral lines. The uncertainties were propagated by dividing through by the median spectrum. We ran {\tt SYSREM} for up to ten consecutive iterations and obtained a residual spectral matrix for each iteration and wavelength segment. An overview of the data reduction procedures, including {\tt SYSREM}, is shown in Fig.~\ref{figure:preprocessing-sysrem}.

%

\section{Detection of the planetary emission lines}
\label{Detection of the planetary emission lines}

\begin{figure*}
        \centering
        \includegraphics[width=\textwidth]{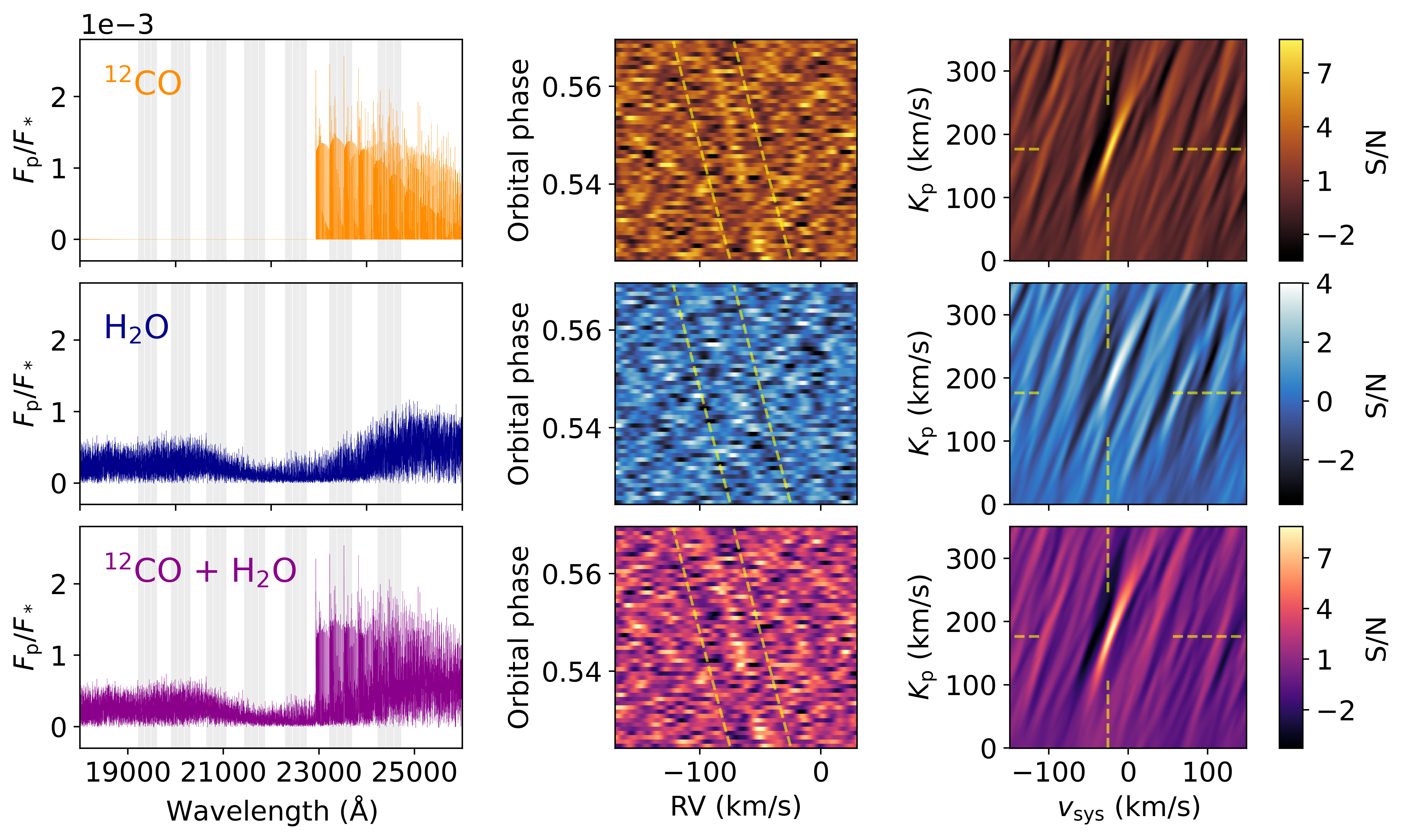}
        \caption{Spectral models, CCF maps, and S/N maps of $^{12}$CO, H$_2$O, and both species together. The \textit{left panels} show the continuum normalized model spectra. Wavelengths covered by the CRIRES$^+$ K2166 setting used in this work are indicated by the gray shaded area. The \textit{middle panels} illustrate the CCF maps. The spectral signal from WASP-178b's atmosphere can be identified as a diagonal trail. The expected RV evolution as a function of orbital phase is indicated by the yellow dashed lines. The \textit{right panels} show the S/N maps. The expected position of the planetary signal is indicated by the yellow dashed lines. We show the CCF maps and S/N maps corresponding to two consecutive runs of \texttt{SYSREM}, which is the number of iterations used in the retrieval in Sect.~\ref{Retrieval of the atmospheric properties}.}
        \label{figure:detections}
\end{figure*}

We used the cross-correlation method \citep[e.g.,][]{Snellen2010, Brogi2012, Birkby2013} to extract the faint spectral emission lines of WASP-178b from the noise-dominated residual spectra. This method consists of calculating the cross-correlation function (CCF) between an exoplanet model spectrum and the residual spectra, thereby mapping the information of the faint planetary spectral lines onto a detectable signal peak. We searched for the spectral signature of CO, H$_2$O, and OH, which are the chemical species with the most prominent emission lines expected in the \mbox{K band} for any reasonable atmospheric composition. In the specific case of CO, the cross-correlation analysis was performed individually on the molecules with the two isotopologues $^{12}$CO and $^{13}$CO. In addition, our search included the emission signal of \ion{Fe}{i}, which has recently been detected with the same CRIRES$^+$ setting as used in this work in the atmosphere of another UHJ, MASCARA-1b \citep{Ramkumar2023}.

\subsection{Model spectra}
\label{Model spectra}

We considered a planetary atmosphere consisting of 81 layers, equispaced in logarithmic pressure from $10^{-8}$\,bar to 1\,bar. Since the thermal conditions in the dayside atmosphere of \mbox{WASP-178b} have not yet been studied in detail, we adopted the $T$-$p$ profile measured for \mbox{KELT-20b/MASCARA-2b} \citep{Yan2022a} to generate the model spectra for a preliminary cross-correlation analysis. Using the $T$-$p$ profile of this planet is a reasonable approximation because its physical properties are close to those of WASP-178b (e.g., planetary radius, equilibrium temperature, orbital period, host star type). By applying the cross-correlation method described in Sect.~\ref{Cross-correlation method}, we obtained a significant detection of the $^{12}$CO and H$_2$O emission lines. We then used the retrieval framework detailed in Sect.~\ref{Retrieval of the atmospheric properties}, which allowed us to determine the best-fit $T$-$p$ profile of WASP-178b's dayside atmosphere. The best-fit atmospheric profile was used to generate the final model spectra, which yielded the cross-correlation analysis results described below.

The $T$-$p$ profile is parametrized by a low pressure point \mbox{($T_1$, $p_1$)} and a high pressure point \mbox{($T_2$, $p_2$)}, with an isothermal atmosphere at pressures below $p_1$ or higher than $p_2$ (Fig.~\ref{figure:VMR-Tp}). The temperature between the two isothermal layers was assumed to change linearly as a function of $\log{p}$. Assuming equilibrium chemistry and the elemental abundances \citep{Asplund2009} retrieved in Sect.~\ref{Retrieval of the atmospheric properties}, we used \texttt{FastChem} \citep{Stock2018}\footnote{\url{https://github.com/exoclime/FastChem}} to obtain the mean molecular weight and volume mixing ratio (VMR) of the chemical species studied (Fig.~\ref{figure:VMR-Tp}). In addition, for each chemical species we needed a \mbox{$T$-$p$-wavelength} grid of its opacities to calculate the corresponding model spectrum. The data used to compute the high-resolution opacities were obtained from the following sources: CO line information from \cite{Li2015}, H$_2$O data from the POKAZATEL line list \citep{Polyansky2018}, OH opacities from the MoLLIST database \citep{Brooke2016}, \ion{Fe}{i} atomic line data from \cite{Kurucz2018}. We ran the radiative transfer code \texttt{petitRADTRANS} \citep{Molliere2019} to generate the high-resolution model spectra of the different chemical species ($R$\,$=$\,$10^6$). Finally, the resulting model spectra were convolved with the rotational broadening profile described in Sect.~\ref{Retrieval of the atmospheric properties}.

\begin{figure*}
        \centering
        \includegraphics[width=\textwidth]{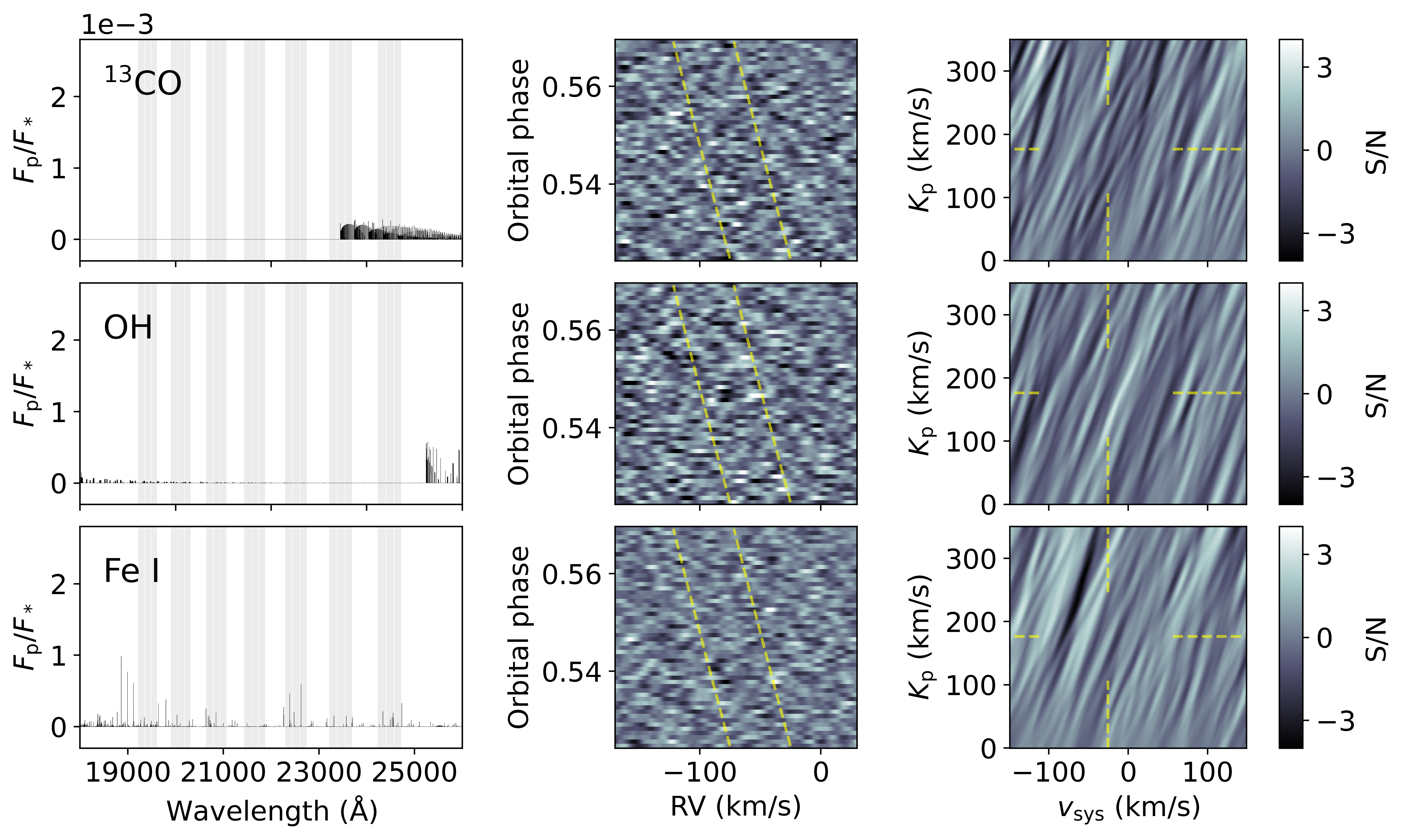}
        \caption{Same as Fig.~\ref{figure:detections}, but for $^{13}$CO, OH, and \ion{Fe}{i}. These species result in non-detections due to the weakness and low number of spectral lines over the investigated wavelength range. The CCF maps and S/N maps are the result of two consecutive {\tt SYSREM} iterations.}
        \label{figure:non-detections}
\end{figure*}

During the pre-processing steps of our analysis, the spectra were normalized to the continuum level. Thus, our model spectra also required continuum normalization. As a first step, we divided each spectral model by the blackbody spectrum of the host star. The resulting planet-to-star flux ratio was then normalized to the planetary continuum and convolved with the instrumental profile, yielding the final model spectra for cross-correlation. The normalized model spectra of all studied chemical species are shown in Figs.~\ref{figure:detections} and \ref{figure:non-detections}.

\subsection{Cross-correlation method}
\label{Cross-correlation method}

We used the cross-correlation technique to study each chemical species independently. The model spectrum was Doppler-shifted to a grid of velocities between \mbox{--800\,km\,s$^{-1}$} and \mbox{+800\,km\,s$^{-1}$}, with a cadence of 1\,km\,s$^{-1}$. We multiplied the shifted model spectrum of each grid point by the uncertainty-weighted residual spectra, resulting in a so-called weighted CCF for each exposure frame. For the residual spectrum number $i$ of the time series, the CCF is thus defined as
\begin{equation}
    \mathrm{CCF}_i\left(\varv\right) = \sum_j \frac{ R_{ij} m_j\left(\varv\right) }{ {\sigma_{ij}}^2 },
\end{equation}
with $R_{ij}$ being the matrix of residual spectra, $\sigma_{ij}$ the corresponding uncertainties, and $m_j$ the spectral model shifted by the Doppler-velocity $\varv$. For each wavelength segment, the CCFs of the spectral time series were arranged in a two-dimensional array. The arrays of the different wavelength segments were then co-added, resulting in a CCF map for each of the A and B spectral subsets. Eventually, the CCF maps of the two nodding positions were combined into a final CCF map for the individual spectral models.

The CCF map was aligned to the rest frame of WASP-178b over a range of different orbital semi-amplitude velocity ($K_\mathrm{p}$) values. For this purpose, we assumed a circular planetary orbit, whose radial velocity is described by
\begin{equation}
    \label{equation:planetary-rest-frame} 
    \varv_\mathrm{p} \left( t \right) = \varv_\mathrm{sys} + \varv_\mathrm{bary} \left( t \right) + K_\mathrm{p} \sin{2\pi\phi \left( t \right)},
\end{equation}
where $\varv_\mathrm{sys}$ is the systemic velocity, $\varv_\mathrm{bary} \left( t \right)$ is the barycentric velocity correction, and $\phi \left( t \right)$ is the orbital phase. Each shifted CCF map was then collapsed into a one-dimensional CCF by averaging over all orbital phases. The one-dimensional CCF of each alignment was further stacked into a two-dimensional array. To quantify the noise level of the array as objectively as possible, we fit the distribution of all CCF values with a Gaussian function \citep[e.g.,][]{Brogi2023}. After dividing the array with the standard deviation of the Gaussian, we obtained a signal-to-noise detection map (S/N map) as a function of $K_\mathrm{p}$ and $\varv_\mathrm{sys}$. 
If the signature of the investigated chemical species is present in the planetary spectrum, a significant detection peak will appear at the expected values of $K_\mathrm{p}$ and $\varv_\mathrm{sys}$ in the S/N map.

\subsection{Cross-correlation results and discussion}
\label{Cross-correlation results and discussion}

\begin{figure}
        \centering
        \includegraphics[width=\columnwidth]{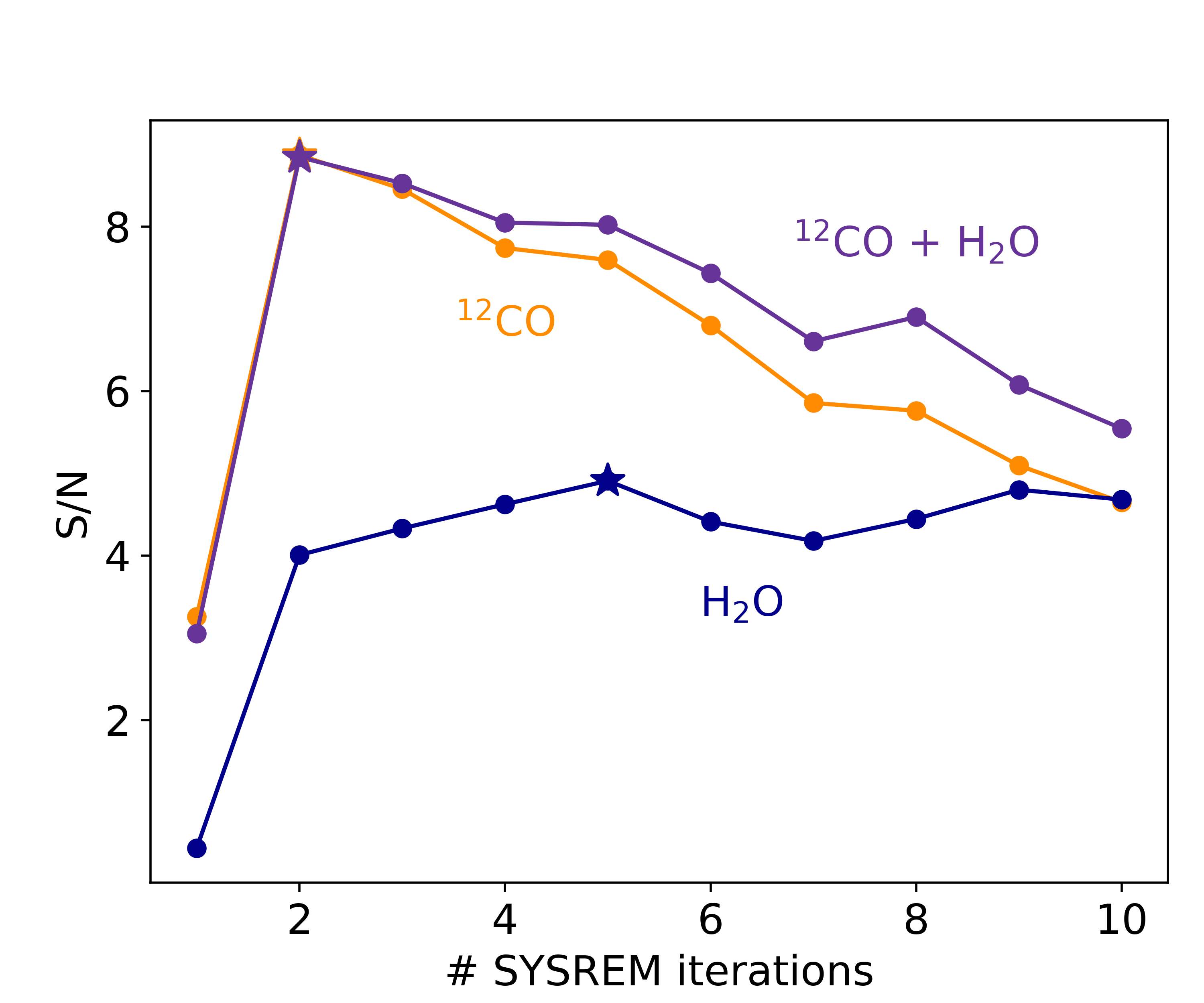}
        \caption{S/N values as a function of \texttt{SYSREM} iterations. We show the detection strengths obtained from cross-correlation with model spectra including the emission lines of $^{12}$CO, H$_2$O, and both species together, respectively. The iteration with the most significant S/N peak is indicated by the star symbol. In our retrieval framework in Sect.~\ref{Retrieval of the atmospheric properties}, we used the iteration number that yields the most prominent S/N detection peak of the species-combined signal, i.e., two consecutive \texttt{SYSREM} iterations.}
        \label{figure:SN-sysrem-iterations}
\end{figure}

We significantly detect spectral emission lines from the dayside atmosphere of WASP-178b. Figure~\ref{figure:detections} shows the CCF maps and S/N maps from the cross-correlation of our data with the $^{12}$CO, H$_2$O, and the species-combined model spectra. The dashed lines indicate the expected location of the planetary signal in the maps as computed from the parameters of \cite{Rodriguez-Martinez2020}. For all {\tt SYSREM} iterations larger than one, both chemical species present a clearly visible CCF trail and a S/N detection peak close to the predicted values of $K_\mathrm{p}$ and $\varv_\mathrm{sys}$. We find the strongest detection of the $^{12}$CO spectral signature with S/N\,=\,8.9 after two {\tt SYSREM} iterations; the most prominent H$_2$O signal is achieved with S/N\,=\,4.9 after five consecutive runs of {\tt SYSREM}. Cross-correlation with a model spectrum including the spectral lines of both chemical species yields a maximum detection after two {\tt SYSREM} iterations with S/N\,=\,8.8. Comparison of this detection strength with the S/N values of the individual species shows that the spectral signature of $^{12}$CO dominates the combined signal. This finding is reasonable considering the increased thermal stability of $^{12}$CO compared to H$_2$O, resulting in higher abundances and stronger spectral lines of this species (Figs.~\ref{figure:VMR-Tp} and \ref{figure:detections}). The detection strengths, signal positions, as well as the noise pattern in the S/N maps of both chemical species show a stable pattern for all {\tt SYSREM} iterations larger than one. We show the evolution of the detection strengths as a function of {\tt SYSREM} iterations in Fig.~\ref{figure:SN-sysrem-iterations}. The detection of spectral lines in emission shape unambiguously proves the presence of an atmospheric temperature inversion in the dayside of WASP-178b, consistent with theoretical predictions on the thermal structure of UHJ atmospheres \citep[e.g.,][]{Hubeny2003, Fortney2008}.

\begin{figure}
        \centering
        \includegraphics[width=\columnwidth]{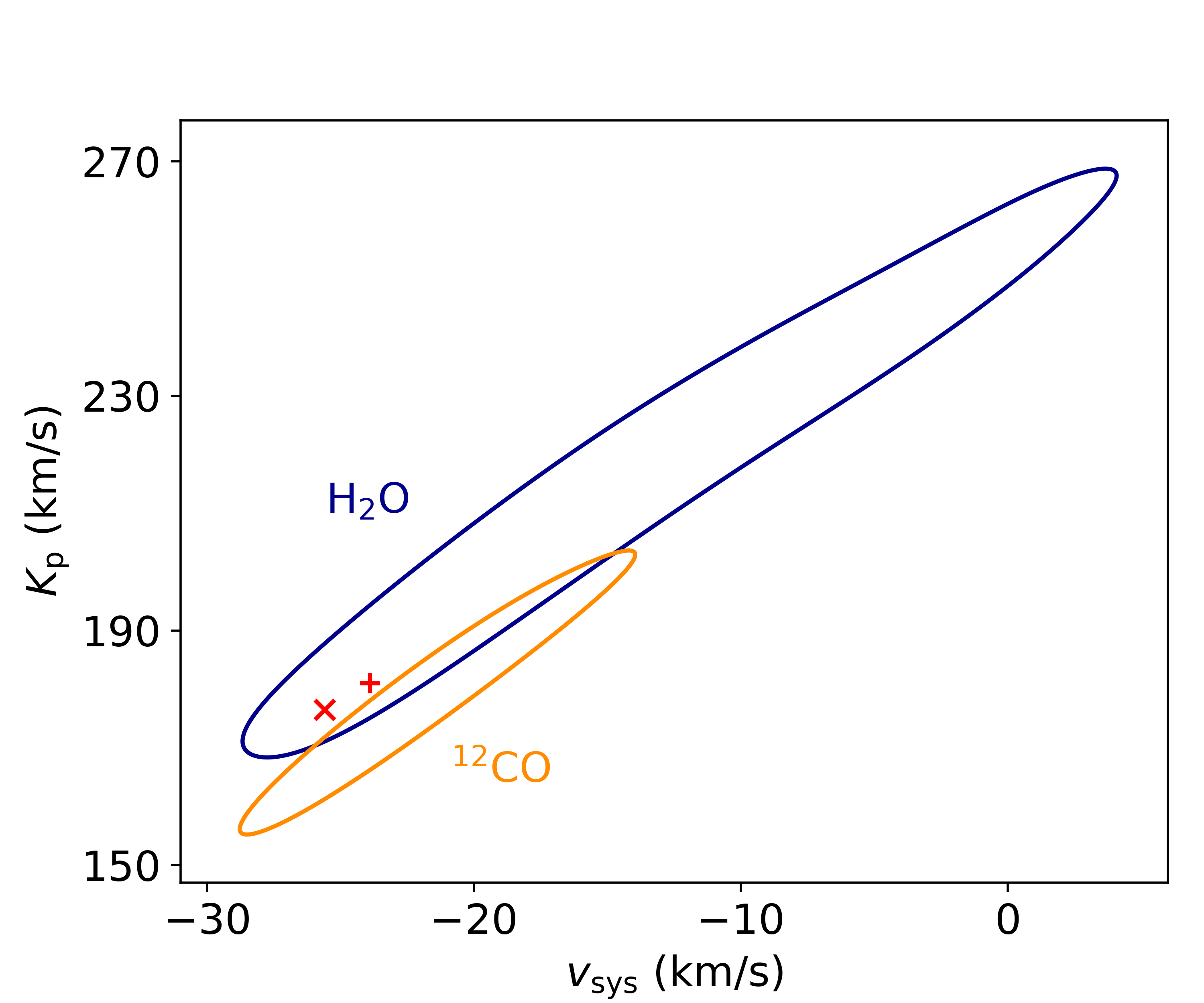}
        \caption{One-sigma regions of $^{12}$CO and H$_2$O detections. The position of the $^{12}$CO signal is inconsistent, that of the H$_2$O signal is consistent with the expected $K_\mathrm{p}$ and $\varv_\mathrm{sys}$ values, whose location is indicated by the red symbols ("\texttt{+}": \citealt{Hellier2019}; "\texttt{$\times$}": \citealt{Rodriguez-Martinez2020}).}
        \label{figure:CO-H2O-1sigma-regions}
\end{figure}

Figure~\ref{figure:CO-H2O-1sigma-regions} illustrates a detailed view of the one-sigma regions around the $^{12}$CO and H$_2$O detection peaks. The signals of the two species show an offset in the order of a few km\,s$^{-1}$ in \mbox{$K_\mathrm{p}$-$\varv_\mathrm{sys}$} space relative to each other, with only a marginal overlap. The different Doppler-shifts prevent a constructive addition of the $^{12}$CO and H$_2$O signals in the S/N map in Fig.~\ref{figure:detections}, explaining the species-combined detection strength being slightly lower than that of $^{12}$CO alone. The $^{12}$CO signal is inconsistent, that of H$_2$O is consistent with the expected orbital parameter values. This is not the first time that signals from different chemical species have been found to have deviating Doppler-shifts. For example, recent observations of the UHJ WASP-18b have revealed a similar effect for the spectral emission lines of CO, H$_2$O, and OH with comparable amplitude in both $K_\mathrm{p}$ and $\varv_\mathrm{sys}$ \citep{Brogi2023}. In addition, a significant offset between the signatures of \ion{Fe}{i} and TiO was identified in the dayside emission spectrum of WASP-33b \citep{Cont2021}. Transmission spectroscopy of gas giant exoplanets has also shown the existence of significant offsets between the detection coordinates in the S/N maps of individual chemical species \citep[e.g.,][]{Kesseli2022, Sanchez-Lopez2022a}. We rule out calibration inaccuracies as a possible cause for the observed offset between the $^{12}$CO and H$_2$O signals, since the precision of our {\tt molecfit} wavelength solution is below the size of a single resolution element of the instrument. Furthermore, the offsets are unlikely to be caused by the line lists used, since previous high-resolution spectroscopy studies have been able to detect $^{12}$CO and H$_2$O simultaneously without detecting a significant Doppler-shift between the two chemical species \citep[e.g.,][]{HolmbergMadhusudhan2022, Ramkumar2023}. In addition, the results of \cite{Gandhi2020} suggest the suitability of the $^{12}$CO and H$_2$O line lists to study exoplanet atmospheres.

Offsets between the signatures of individual chemical species in $K_\mathrm{p}$-$\varv_\mathrm{sys}$ space may indicate a difference in the local ``slope'' of the Keplerian radial velocity curve, caused by atmospheric circulation and chemical inhomogeneity effects \citep{Brogi2023}. The concentration of $^{12}$CO decreases significantly slower than that of H$_2$O towards lower pressure values (Fig.~\ref{figure:VMR-Tp}). Consequently, the emission lines of the two species are expected to probe different pressure regimes in WASP-178b's atmosphere. Doppler-offsets between the individual chemical species may therefore be caused by differences in the dynamics of the respective atmospheric layers. In addition, chemical inhomogeneities across WASP-178b's surface, in combination with the planetary rotation, have the potential to add an individual Doppler-shift to the signature of each chemical species. 
While the existence of an offset between the measured $^{12}$CO peak and the expected $K_\mathrm{p}$-$\varv_\mathrm{sys}$ coordinates is confidently detected, we emphasize that caution is required in interpreting the position of the weaker H$_2$O signal in the S/N map. The one-sigma region of the species extends over a wide range of orbital parameter values, making it difficult to draw precise conclusions about the Doppler-shift of the signal. In addition, the relatively low signal strength raises the possibility that the observed position in $K_\mathrm{p}$-$\varv_\mathrm{sys}$ space could be affected by distortion of the H$_2$O peak due to the intrinsic noise pattern of the S/N map. Future observations will provide a stronger H$_2$O signal, improving the ability to study the three-dimensional structure and dynamics of WASP-178b's atmosphere.

\begin{figure*}
        \centering
        \vspace{0.3cm}
        \includegraphics[width=\textwidth]{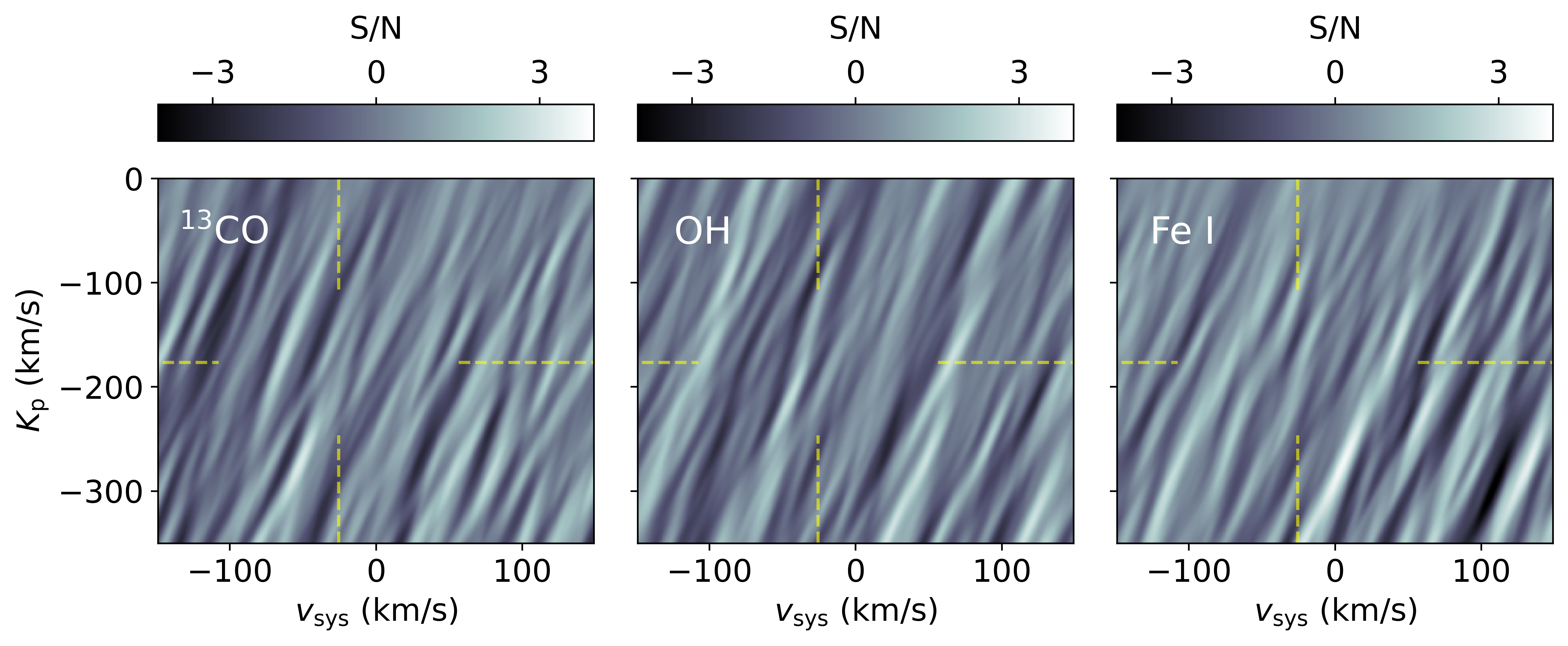}
        \vspace{0.1cm}        
        \caption{S/N maps of injection-recovery test. The yellow dashed lines indicate the coordinates of the injected signal. The injected signatures of $^{13}$CO, OH, and \ion{Fe}{i} could not be identified, indicating that even if present in the planetary atmosphere, these species are undetectable.}
        \label{figure:injection-recovery}
\end{figure*}

Figure~\ref{figure:non-detections} shows the S/N maps of $^{13}$CO, OH, and \ion{Fe}{i}. No significant detections are found in the S/N maps of the three chemical species. We conducted an injection-recovery test to validate our findings. In a first step, we Doppler-shifted the convolved model spectra used to attempt the detection of the chemical species with the reversed $K_\mathrm{p}$ value of --176.5\,km\,s$^{-1}$ \citep{Merritt2020, Cont2022b}. The shifted spectral models were then injected into the raw data, and the pre-processing and cross-correlation analyses were performed as described in Sects.~\ref{Data reduction} and \ref{Detection of the planetary emission lines}. Doppler-shifting the injected signal with the negative $K_\mathrm{p}$ of WASP-178b serves to avoid interference with potentially undetected planetary emission lines of the investigated chemical species. None of the injected signals were recovered (Fig.~\ref{figure:injection-recovery}). The nondetections of the chemical species $^{13}$CO, OH, and \ion{Fe}{i} are therefore most likely due to the limited number and low intensity of their spectral emission lines in the wavelength range studied, rather than to their absence in the planetary atmosphere. In fact, OH emission lines were identified in the spectra of gas giant exoplanets other than WASP-178b, but blueward of the wavelength range analyzed in this work \citep[e.g.,][]{Nugroho2021, Brogi2023}. In addition, \ion{Fe}{i} lines have commonly been identified at visible wavelengths in the spectra of UHJs \citep[e.g.,][]{Pino2022, Yan2022a}. The emission signature of $^{13}$CO has also been detected at near-infrared wavelengths by \cite{Zhang2021}, albeit by medium-resolution spectroscopy for a gas giant at a significantly larger orbital separation compared to UHJs.

%

\section{Retrieval of the atmospheric properties}
\label{Retrieval of the atmospheric properties}

Retrieval frameworks allow the derivation of statistical constraints on the physical and chemical parameters of exoplanet atmospheres by comparing a parameterized model spectrum with observational data. In this study, we adopt the approach previously used by \cite{Yan2023} and \cite{Lesjak2023} to quantitatively assess the atmospheric conditions of WASP-178b. This framework is an evolution of the retrieval method introduced by \cite{Yan2020}, inspired by \cite{Brogi-Line2019}, \cite{Shulyak2019}, and \cite{Gibson2020}, and successfully used in previous work on UHJs \citep{Yan2022a, Cont2022b, Borsa2022a}. 
The main advancement in this retrieval is the integration of a method that accounts for potential distortions in the spectral data introduced by {\tt SYSREM}.

\subsection{Retrieval framework}
\label{Retrieval framework}

\subsubsection{High-resolution retrieval}
\label{Retrieval framework on CRIRES$^+$ data}

\begin{figure}
        \centering
        \includegraphics[width=\columnwidth]{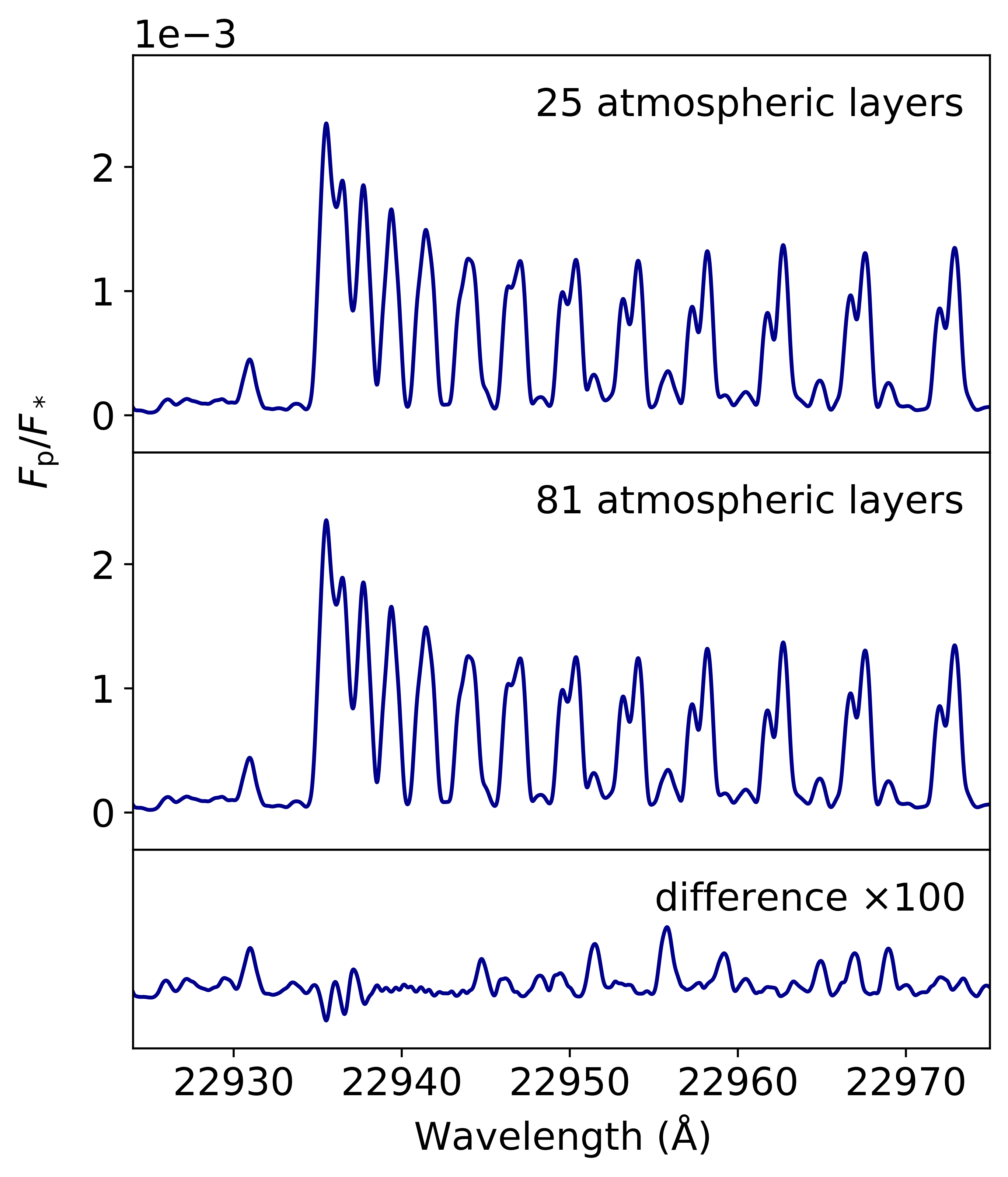}
        \caption{Comparison between model spectra computed with 25 (\textit{top panel}) and 81 (\textit{middle panel}) atmospheric layers. The model spectra include the opacities of the detected chemical species $^{12}$CO and H$_2$O, as well as those of H$^-$. The diﬀerence between the models is insignificant (\textit{bottom panel}), implying that the lower number of atmospheric layers does not significantly affect our retrieval results.}
        \label{figure:compare_12CO_H2O_combined_25vs81_layers}
\end{figure}

We used the radiative transfer code {\tt petitRADTRANS} \citep{Molliere2019} to forward model the emission spectrum of \mbox{WASP-178b} ($R$\,=\,$10^6$). Our radiative transfer calculations included the opacities of the detected chemical species $^{12}$CO and H$_2$O, as well as those of H$^-$, which has been suggested as an important continuum source capable of muting emission features in the spectra of UHJs \citep[e.g.,][]{Arcangeli2018, Lothringer2018}. The use of {\tt petitRADTRANS} requires the definition of the thermal and chemical properties of the investigated planetary atmosphere. For this purpose, we modeled the atmosphere of \mbox{WASP-178b} with 25 layers uniformly distributed on a logarithmic scale between $10^{-8}$\,bar and 1\,bar. 
The reduced number of atmospheric layers compared to those used to generate the spectral models in Sect.~\ref{Model spectra} allows us to speed up our radiative transfer calculations. We have verified that the number of layers used has a negligible impact on the accuracy of the calculated model spectra (Fig.~\ref{figure:compare_12CO_H2O_combined_25vs81_layers}). 
As outlined in Sect.~\ref{Model spectra}, a two-point \mbox{$T$-$p$} parametrization was used to describe the thermal profile of the atmosphere. Instead of using the high pressure point $p_2$ directly, we define it as $\log{p_2} = \log{p_1} + \mathrm{d}p$. We limited $\mathrm{d}p$ to positive values to ensure that $p_2$ has a value higher than $p_1$. The chemical equilibrium code \texttt{FastChem} \citep{Stock2018} was employed to calculate the VMRs as a function of the atmospheric metallicity ([M/H]) and C/O ratio. To account for the non-negligible effect of WASP-178b's rotation, we convolved the {\tt petitRADTRANS} forward model with a broadening profile parametrized by the planetary equatorial rotation velocity $\varv_\mathrm{eq}$. The analytical expression of the broadening profile is given in Eq.~3 of \cite{Diaz2011} and has been commonly used to study stellar rotation. We assumed a limb darkening coefficient of $\epsilon$\,=\,1, corresponding to a flux contribution tending to zero towards the planetary limb. In addition, the inclination angle of WASP-178b's equator was set to the same value as the orbital inclination angle of the planet, under the hypothesis of a tidally locked rotation. The model spectrum was further converted to the planet-to-star flux ratio and convolved with the instrumental profile as described in Sect.~\ref{Model spectra}. 

We Doppler-shifted the model spectrum to the radial velocity of each exposure frame of our spectral time series. The shift was calculated using Eq.~\ref{equation:planetary-rest-frame}, thus depending on the values of $K_\mathrm{p}$ and $\varv_\mathrm{sys}$. We interpolated each Doppler-shifted model spectrum to the wavelength solution of the observational data points and arranged the resulting one-dimensional vectors in a two-dimensional array. This procedure yielded a two-dimensional forward model matrix, with the same shape as the residual spectral matrix. 

The presence of distortion effects on the spectral emission lines introduced by {\tt SYSREM} was incorporated into the forward model matrix. To this end, we adopted the filtering method proposed by \cite{Gibson2022}, which avoids the time-consuming application of {\tt SYSREM} to the two-dimensional forward model at each sampling step of the retrieval. This technique consists of applying the filter 
\begin{equation}
    \mathbf{F} = \mathbf{A}( \mathbf{\Lambda} \mathbf{A})^\dagger ( \mathbf{\Lambda} \mathbf{M}_\mathrm{unfiltered})
\end{equation}
to the unfiltered forward model matrix $\mathbf{M}_\mathrm{unfiltered}$. The matrix $\mathbf{A}$ is defined in Sect.~\ref{Removal of telluric and stellar lines}, $\mathbf{\Lambda}$ is a diagonal matrix containing the inverse of the mean uncertainties in wavelength direction, and $\mathbf{X}^\dagger = (\mathbf{X}^T\mathbf{X})^{-1}\mathbf{X}^T$ denotes the Moore-Penrose inverse of a matrix $\mathbf{X}$. We pre-calculated and applied the filter to each model during the retrieval, resulting in a filtered model matrix $\mathbf{M} = \mathbf{M}_\mathrm{unfiltered} - \mathbf{F}$, which was compared to the data. 

A standard Gaussian log likelihood function
\begin{equation}
    \ln{L} = -\frac{1}{2}\sum_{i,j} \left[ \frac{\left(R_{ij} - M_{ij}\right)^2}{\left(\beta \sigma_{ij}\right)^2} + \ln{2 \pi \left(\beta \sigma_{ij}\right)^2} \right]
\end{equation}
was defined to compare each generated forward model with the residual spectra. In this expression, $R_{ij}$ and $M_{ij}$ denote the individual elements of the residual spectral matrix and the filtered forward model matrix, respectively. The uncertainties of the residual spectra are represented by $\sigma_{ij}$, while $\beta$ acts as a scaling factor to correct for possible over- or underestimation of the uncertainties. We used the residual spectral matrix that gives the most prominent S/N detection peak, which was obtained after two consecutive {\tt SYSREM} iterations (see Fig.~\ref{figure:SN-sysrem-iterations}). To confirm that our choice of the iteration number has no significant effect on the retrieval result, we repeated our calculations with the spectral matrices from {\tt SYSREM} iterations other than the optimal number of two consecutive runs, for which we did not find any significant differences. For each wavelength segment and nodding position, the log likelihood function was calculated independently, and the resulting values were summed to obtain the combined log likelihood function of all data. To estimate the model parameters, we evaluated the combined log likelihood function by Markov Chain Monte Carlo (MCMC) sampling using the \texttt{emcee} software package \citep{Foreman-Mackey2013}.

In summary, our high-resolution retrieval framework includes the following free parameters: the temperature profile parameters $T_1$, $p_1$, $T_2$, d$p$; the chemical properties represented by [M/H] and the C/O ratio; the equatorial rotation velocity $\varv_\mathrm{eq}$; the velocity parameters $K_\mathrm{p}$ and $\varv_\mathrm{sys}$; the noise scaling parameter $\beta$. For each free parameter, we used 32 walkers with 15\,000 steps in the sampling.

\subsubsection{Joint retrieval with photometric data}
\label{Retrieval including TESS and CHEOPS data}

High-resolution spectroscopy observations lack the information from the spectral continuum level, which is lost when correcting for the telluric and stellar lines with {\tt SYSREM}. Therefore, only the relative strengths of the spectral lines with respect to the local continuum can be measured, reducing the reliability of retrievals when relying solely on high-resolution spectroscopy. On the other hand, space-based observations can measure absolute flux levels that encode the spectral continuum information of exoplanet atmospheres. An analysis that combines our high-resolution CRIRES$^+$ data with space-based measurements will therefore improve the robustness of our retrieval results. 

Recently, \cite{Pagano2023} measured the secondary eclipse depth of WASP-178b using space-based photometry with TESS and CHEOPS, obtaining eclipse depth values of 70$\pm$20\,ppm and 70$\pm$40\,ppm, respectively. We incorporated these two photometric data points into our retrieval framework using an approach similar to that proposed by \cite{Yan2022a}, which we complemented by including the reflection of stellar light by the planetary atmosphere. Given the photometric passband $\eta \left(\lambda\right)$, this approach models the eclipse depth as
\begin{equation}
\label{equation:eclipse-depth}
    \delta_\mathrm{ecl} = A_\mathrm{g} \left( \frac{R_\mathrm{p}}{a} \right)^2 + \left( \frac{R_\mathrm{p}}{R_*} \right)^2 \frac{\int \eta\left(\lambda\right) F_\mathrm{p}\left(\lambda\right) \, d\lambda}{\int \eta\left(\lambda\right) F_*\left(\lambda\right) \, d\lambda},
\end{equation}
where $A_\mathrm{g}$ is the geometric albedo, $a$ is the orbital separation of the planet-star system, and $R_\mathrm{p}$ and $R_*$ are the planetary and stellar radii. Additionally, $F_\mathrm{p}\left(\lambda\right)$ and $F_*\left(\lambda\right)$ represent the fluxes originating from the planetary and stellar atmospheres per unit area and per unit wavelength. The first term of this expression accounts for the contribution of the stellar light reflected by the planetary atmosphere; the second term represents the thermal emission contribution to the planet-to-star flux ratio integrated over the instrumental passband. 

We used {\tt petitRADTRANS} to calculate the thermal emission component of Eq.~\ref{equation:eclipse-depth} over the wavelength range of TESS and CHEOPS. The photometric bandpasses of both instruments cover a significant fraction of the visible wavelength range where strong emission lines of atomic metals, their oxides, and hydrides are expected. In our radiative transfer calculations, we included the opacities of Fe, Ti, Na, K, Ca, Mg, TiO, VO, FeH, CrH, and CaH, which represent the metal species with the most prominent emission features at visible wavelengths for any reasonable atmospheric composition \citep[e.g.,][]{Hubeny2003, Fortney2008, Lothringer2019}. Additionally, we accounted for the contribution from the continuum opacity source H$^-$. We provided {\tt petitRADTRANS} with the same thermal and chemical parametrizations as detailed in Sect.~\ref{Retrieval framework on CRIRES$^+$ data}. The resulting emission flux of each instrument was integrated over its respective bandpass using the {\tt rebin-give-width} tool of {\tt petitRADTRANS}. We integrated over the range 0.60 to 1.00\,$\mu$m for TESS, while for CHEOPS the integration was performed over the range 0.35 to 1.10\,$\mu$m. Since the model spectrum was integrated over the relatively wide TESS and CHEOPS bandpasses, we used the {\tt petitRADTRANS} low-resolution mode to speed up the radiative transfer calculations. 
The opacities used in the low-resolution mode require less dense sampling in wavelength space than those used to generate the high-resolution CRIRES$^+$ model spectra, and are available in the \texttt{petitRADTRANS} opacity database \footnote{\url{https://petitradtrans.readthedocs.io/en/latest/content/available_opacities.html}}.

We then defined the log likelihood function for the photometric data. For a specific eclipse depth measurement, this function is given by
\begin{equation}
    \ln{L_\mathrm{ecl}} = -\frac{1}{2} \left[ \frac{(R_\mathrm{ecl} - \delta_\mathrm{ecl})^2}{\sigma_\mathrm{ecl}^2} + \ln{2 \pi \sigma_\mathrm{ecl}^2} \right],
\end{equation}
where $R_\mathrm{ecl}$ is the photometric eclipse depth measurement, $\delta_\mathrm{ecl}$ is the model of the eclipse depth, and $\sigma_\mathrm{ecl}$ is the measurement uncertainty. We calculated the log likelihood functions for both the TESS and CHEOPS data points. We then added these functions to the combined log likelihood function of the high-resolution data described in Sect.~\ref{Retrieval framework on CRIRES$^+$ data}. The resulting final log likelihood function incorporates all the information from the high-resolution and photometric data. We performed the joint retrieval on the CRIRES$^+$, TESS, and CHEOPS data by evaluating the final log likelihood function via MCMC sampling. In addition to the free parameters outlined in Sect.~\ref{Retrieval framework on CRIRES$^+$ data}, the retrieval incorporates the geometric albedo $A_\mathrm{g}$. We set $A_\mathrm{g}$ as gray albedo, which assumes wavelength independence. For each free parameter, 32 walkers with 15\,000 steps each were used in the MCMC sampling process.

\subsection{Results and discussion}
\label{Results and discussion}


\begin{table*}
        \caption{Results of atmospheric retrievals on WASP-178b.
        }             
        \label{tab:retrieval-results}      
        \centering                          
        \renewcommand{\arraystretch}{1.5} 
        \begin{threeparttable}
                \begin{tabular}{l c c c c}        
                        \hline\hline                 
                        \noalign{\smallskip}
                        \multirow{2}*{Parameter} & High-resolution retrieval & Joint retrieval with photometric data  & \multirow{2}*{Prior} & \multirow{2}*{Unit}  \\     
                                  & (CRIRES$^+$ only)         & (CRIRES$^+$ + TESS + CHEOPS) &  &   \\                            
                        \noalign{\smallskip}
                        \hline                       
                        \noalign{\smallskip}
                        $T_1$                  & $>3661$                      & $>3182$                     & (1000, 7000)     &    K \\ 
                        $\log{p_1}$            & $-4.62_{-1.88}^{+1.56}$      & $-5.46_{-1.51}^{+1.38}$     & (-8, 0)          &    log bar\\ 
                        $T_2$                  & $<2756$                      & $<2038$                     & (100, 7000)    &    K \\ 
                        $\mathrm{d}p$          & $4.46_{-2.19}^{+2.19}$       & $2.94_{-1.58}^{+2.12}$      & (0, 8)         &    log bar\\    
                        $\log{p_2}$            & $-0.33_{-1.53}^{+1.64}$      & $-2.52_{-1.06}^{+1.64}$     & \ldots          &    log bar\\    
                        $\mathrm{[M/H]}$       & $0.40_{-0.99}^{+1.81}$       & $1.39_{-1.58}^{+1.64}$      & (-5, 5)         &    dex \\ 
                        $\mathrm{C/O}$         & $<0.84$                      & $0.59_{-0.23}^{+0.20}$      & (0, 3)           &    \ldots\\ 
                        $\varv_\mathrm{eq}$    & $5.65_{-1.13}^{+1.17}$       & $6.74_{-1.03}^{+1.17}$      & (0, 15)          &    km\,s$^{-1}$ \\ 
                        $A_\mathrm{g}$         & \ldots                       & $<0.23$                     & (0, 1)            &    \ldots\\        
                        $K_\mathrm{p}$         & $174.46_{-6.44}^{+7.22}$     & $176.41_{-7.04}^{+8.28}$    & (100, 250)       &    km\,s$^{-1}$\\ 
                        $\varv_\mathrm{sys}$   & $-23.26_{-1.89}^{+2.03}$     & $-22.78_{-2.08}^{+2.37}$    & (-50, 30)        &    km\,s$^{-1}$ \\ 
                        $\beta$                & $1.0635_{-0.0005}^{+0.0006}$ & $1.0636 \pm 0.0005$         & (0, 3)           &    \ldots\\     
                        \noalign{\smallskip}
                        \hline                                   
                \end{tabular}
        \end{threeparttable}      
\end{table*}

\subsubsection{High-resolution retrieval}
\label{Results from CRIRES$^+$ data and discussion}

In a first step, we applied the retrieval only to the high-resolution spectroscopy data obtained from CRIRES$^+$. The corner plot in Fig.~\ref{figure:posterior-crires} shows the resulting posterior distributions along with the correlations of the atmospheric parameters. A summary of the best-fit retrieval parameters is provided in Table~\ref{tab:retrieval-results}. The noise scaling term $\beta$ is close to one, indicating an appropriate estimation of the uncertainties.

The retrieval confirms the presence of the inverted atmospheric $T$-$p$ profile detected in Sect.~\ref{Detection of the planetary emission lines}. Figure~\ref{figure:posterior-crires}a shows examples of the $T$-$p$ proﬁles sampled by the MCMC analysis, and Fig.~\ref{figure:posterior-crires}b illustrates the median temperature curve. The thermal inversion layer spans the pressure range from approximately $10^{-5}$\,bar to 1\,bar, with temperatures of $T_1>3700$\,K and \mbox{$T_2<2800$\,K} in the upper and lower planetary atmosphere, respectively. The strength of the atmospheric temperature inversion is consistent with both theoretical predictions \citep{Hubeny2003, Fortney2008, Arcangeli2018} and previous observational studies of UHJs with system architectures similar to WASP-178b \citep[e.g.,][]{Yan2020, Yan2022a}.

Although we are able to infer the existence of a strong thermal inversion, our retrieval is not able to precisely capture the physical conditions in the uppermost layers of WASP-178b's atmosphere. This can be seen when considering the posterior distribution of the temperature parameter $T_1$ in Fig.~\ref{figure:posterior-crires}, which parametrizes the upper boundary of the thermal inversion layer. In particular, at high values of $T_1$, the posterior probability distribution of the parameter tends to have a flat pattern. In the context of Bayesian retrieval frameworks, such a flat posterior distribution generally indicates a lack of information from the observational data in the specific parameter range. The presence of little spectral information from the upper planetary atmosphere is indeed expected, since $^{12}$CO and H$_2$O generally undergo strong dissociation with decreasing atmospheric pressure. Consequently, the spectral emission signature of $^{12}$CO and H$_2$O will originate primarily from atmospheric layers where higher pressures allow significant abundances of the two molecular species to persist.

Similar to $T_1$, the retrieval on the CRIRES$^+$ data can barely constrain the temperature parameter $T_2$, which describes the lower boundary of the thermal inversion layer. The posterior probability distribution of $T_2$ flattens towards lower values, suggesting that limited information can be recovered in this temperature range. We attribute the poor constraints on $T_2$ to the absence of information from the spectral continuum level, which encodes the physical properties at the bottom of the thermal inversion layer but was removed from the data during the correction for systematic effects in Sect.~\ref{Removal of telluric and stellar lines}. The absence of information on the spectral continuum level also explains the degeneracy between $T_2$ and $p_2$, which leads to large uncertainties in the value of $p_2$. Given the weak constraints on $T_2$ and $p_2$, several of the sampled thermal inversions in Fig.~\ref{figure:posterior-crires}a extend to temperatures below 1000\,K and pressures well above 1\,bar. These temperature values are not plausible given the extreme thermal conditions in UHJ atmospheres, while the elevated pressure values correspond to atmospheric layers inaccessible to spectral observations. In summary, we are unable to accurately measure both the upper and lower boundaries of the thermal inversion layer using high-resolution CRIRES$^+$ data alone.

Our retrieval constrains the atmospheric metallicity to [M/H]\,=\,$0.40_{-0.99}^{+1.81}$\,dex, which is in agreement with the host star value of $-0.06_{-0.34}^{+0.30}$\,dex \citep{Rodriguez-Martinez2020}. In addition, we derived an upper limit for the atmospheric C/O ratio of 0.84, which is consistent with the solar value of 0.55. A degeneracy of the measured [M/H] with the pressure parameters $p_1$ and $p_2$, as well as between the C/O ratio and $p_2$, is indicated by the diagonal distribution patterns in the correlation plots of Fig.~\ref{figure:posterior-crires}. 
Under the hypothesis of equilibrium chemistry, the concentrations of $^{12}$CO and H$_2$O decrease in the upper planetary atmosphere due to the growing influence of molecular dissociation towards lower pressures. The effect of the reduced molecular concentrations on the forward model spectrum can be counterbalanced by increasing the [M/H] value, explaining the degeneracy between the pressure parameters and [M/H]. 
In addition, the concentration of H$_2$O decreases faster than that of $^{12}$CO in the low pressure regime. The influence of the lower H$_2$O concentration on the spectral forward model can be compensated by decreasing the C/O ratio, resulting in the observed degeneracy with the pressure parameter. 
Consequently, uncertainties in the retrieved $T$-$p$ profile have the potential to critically affect the [M/H] and C/O values. More precise constraints on the $T$-$p$ profile will be derived in Sect.~\ref{Joint retrieval with photometric data}, improving the reliability of the parameters that describe the atmospheric chemistry. 

High-resolution spectroscopy retrievals in the literature typically assume constant VMRs, rather than allowing abundances to vary with atmospheric altitude \citep[e.g.,][]{Gandhi2023, Yan2023}. To better capture the chemical conditions as a function of atmospheric pressure, equilibrium chemistry calculations can be incorporated into retrieval frameworks \citep[e.g.,][]{MadhusudhanSeager2009, Line2012, Kitzmann2020}. We note that the equilibrium chemistry approximation is not generally valid over the entire pressure and temperature range of exoplanet atmospheres, as chemical disequilibrium effects can alter the VMRs of different chemical species \citep[e.g.,][]{Shulyak2020}. However, theoretical work has also shown that in the specific case of temperatures in the UHJ regime, the assumption of equilibrium chemistry may be a valid approximation \citep{Kitzmann2018}.

\begin{figure*}
        \centering
        \includegraphics[width=\textwidth]{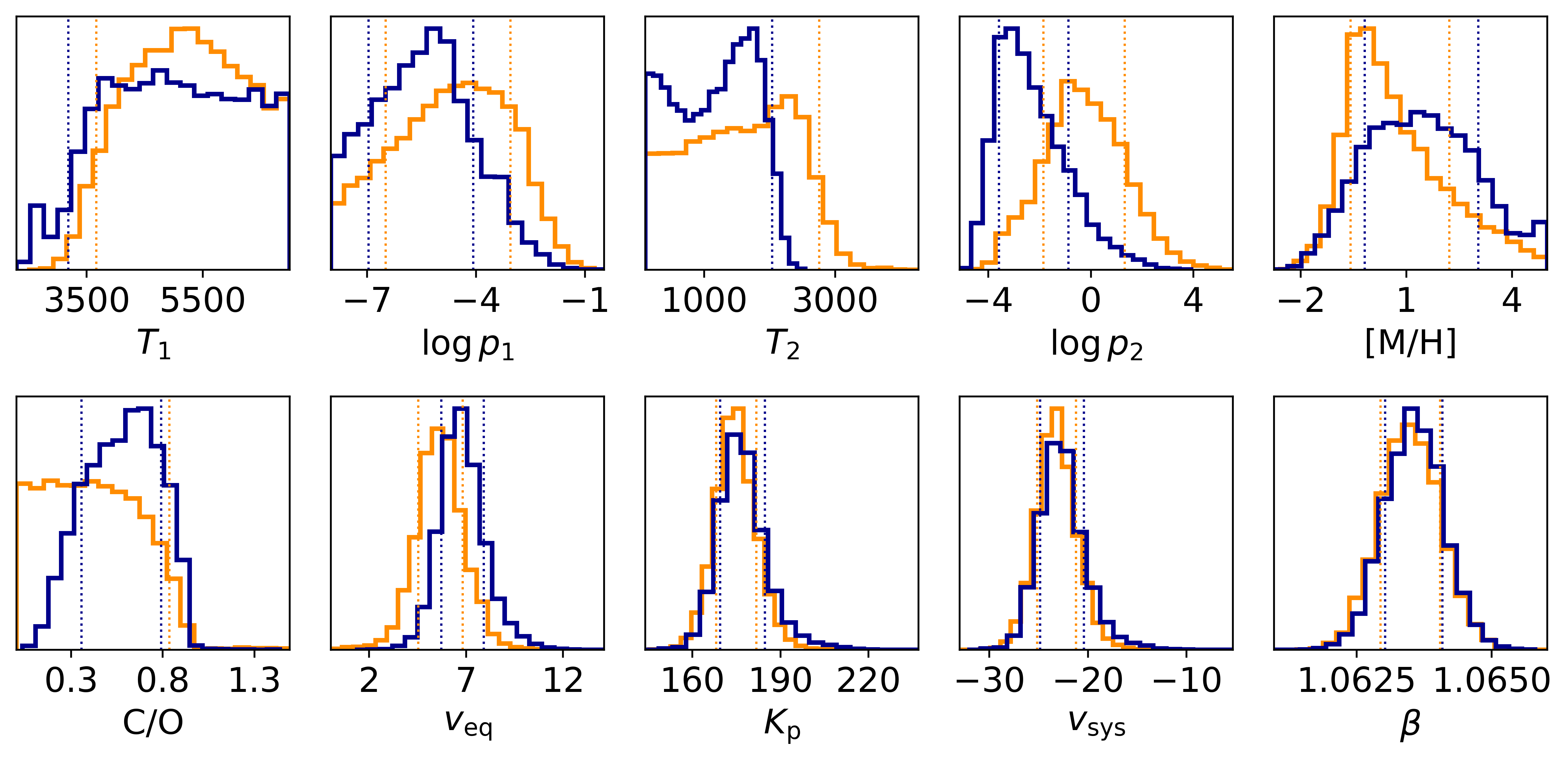}
        \caption{Comparison between the posterior distributions of the high-resolution retrieval (orange; also reported in Fig.~\ref{figure:posterior-crires}) and the joint retrieval with photometric data (blue; also reported in Fig.~\ref{figure:posterior-crires-tess-cheops}). The parameters of both retrievals agree within the 1$\sigma$ confidence intervals for the bounded parameters and the 2$\sigma$ intervals for the upper and lower limits (shown as dashed vertical lines; the median is omitted for clarity). The most pronounced discrepancies are found in the parameter distributions describing the low-altitude atmosphere and the chemical conditions, i.e., $T_2$, $p_2$, [M/H], and C/O.}
        \label{figure:overplot-histograms}
\end{figure*}

In contrast to the somewhat weak constraints on the thermal structure and chemical conditions, our retrieval is able to obtain tight confidence intervals for the planetary parameters encoded in the Doppler-shifts and the width of the spectral emission lines. We retrieve the orbital and systemic velocity parameters with high precision as $K_\mathrm{p}$\,=\,$174.46_{-6.44}^{+7.22}$\,km\,s$^{-1}$ and $\varv_\mathrm{sys}$\,=\,$-23.26_{-1.89}^{+2.03}$\,km\,s$^{-1}$. These velocity values are in line with the location of the S/N detection peak obtained via cross-correlation with the combined model of $^{12}$CO and H$_2$O in Sect.~\ref{Detection of the planetary emission lines}. The derived value of $\varv_\mathrm{sys}$ is in agreement with the measurement reported by \cite{Hellier2019}, while a slight blue-shift is found with respect to the result of \cite{Rodriguez-Martinez2020}. 
Such a blue-shift can generally indicate a net flow of atmospheric gas towards the observer. This phenomenon has allowed the identification of day-to-nightside winds in the atmospheres of several exoplanets via transmission spectroscopy \citep[e.g.,][]{Snellen2010, AlonsoFloriano2019}. Our emission spectroscopy observations, however, correspond to orbital phases close to the secondary eclipse. In this observation geometry, a blue-shifted spectral signal would result from a net material flow from the nightside to the dayside, a dynamical regime not predicted by global circulation models of UHJs \citep[e.g.,][]{Showman2013, TanKomacek2019}. Therefore, and given the agreement between our $\varv_\mathrm{sys}$ value and that of \cite{Hellier2019}, we suggest that the small offset from the \cite{Rodriguez-Martinez2020} measurement is unlikely caused by global winds in WASP-178b's atmosphere. 
Overall, the inconsistency between the results of the two detection papers of \mbox{WASP-178b} makes it difficult to interpret our $\varv_\mathrm{sys}$ measurement in absolute terms. However, the offset does not affect our interpretation of the relative Doppler-shift between the $^{12}$CO and H$_2$O emission lines discussed in Sect.~\ref{Detection of the planetary emission lines}.

In addition, our retrieval reveals the presence of significant spectral line broadening. Assuming that broadening of the spectral emission lines is mainly caused by the planetary rotation, we derive an equatorial rotation velocity of $\varv_\mathrm{eq}$\,=\,$5.65_{-1.13}^{+1.17}$\,km\,s$^{-1}$. We exclude that thermal or pressure broadening cause the increased width of the spectral line profile, since these effects are already included in the forward model via the opacities of the radiative transfer calculation. We also rule out that smearing of the spectral signal over multiple pixels due to the wavelength shift during the observations causes the broadened line profile. The 240\,s exposure time of each spectrum corresponds to a smearing slightly less than 1\,km\,s$^{-1}$, which is significantly below the size of a single CRIRES$^+$ resolution element. The derived $\varv_\mathrm{eq}$ is significantly higher than that expected from a tidally locked rotation of WASP-178b, which would yield a value of $\sim$\,3\,\,km\,s$^{-1}$. This finding indicates the presence of strong atmospheric dynamics, most likely global-scale winds flowing parallel to the equator in the same direction as the planetary rotation. Indeed, global circulation models predict the existence of super-rotating winds in the atmospheres of gas giant exoplanets \citep[e.g.,][]{Showman2009, Heng2011, Beltz2021}. 

In principle, the signature of a super-rotating atmosphere should be observable either as a Doppler-shift \citep{Seidel2023} or as a bi- or trimodal profile \citep{Nortmann2024} of the spectral lines during primary eclipses. Recent transmission spectroscopy observations of WASP-178b at visible wavelengths with ESPRESSO have revealed the absence of such a signature in the absorption lines of atomic metals \citep{Damasceno2024}. Localized zonal winds on the dayside of WASP-178b that dissipate before reaching the terminator could provide an explanation for both the ESPRESSO transmission and CRIRES$^+$ emission spectroscopy observations. Such an atmospheric flow pattern has recently been proposed by \cite{Seidel2023} to explain the absence of a jet stream near the terminators of the UHJ WASP-121b. Alternatively, a bi- or trimodal line profile could be present, but not resolved in the ESPRESSO transmission observations. Future transmission and emission spectroscopy observations over an extended wavelength range and orbital phase interval are needed to improve our understanding of the atmospheric circulation in WASP-178b's dayside.

\subsubsection{Joint retrieval with photometric data}
\label{Joint retrieval with photometric data}

In a next step, we performed the retrieval that includes both the high-resolution spectroscopy data from CRIRES$^+$ and the photometric measurements from TESS and CHEOPS. The corner plot in Fig.~\ref{figure:posterior-crires-tess-cheops} illustrates the resulting posterior distributions and the correlations between the atmospheric parameters. A comparison with the posterior probability distributions of the high-resolution retrieval is shown in Fig.~\ref{figure:overplot-histograms}. The best-fit parameters are summarized in Table~\ref{tab:retrieval-results}. All retrieval parameters except the newly introduced geometric albedo $A_\mathrm{g}$ agree with the results of the high-resolution retrieval within the uncertainty intervals. 

The atmospheric $T$-$p$ profiles obtained from the MCMC sampling process are shown in Fig.~\ref{figure:posterior-crires-tess-cheops}a, while the median temperature profile is reported in Fig.~\ref{figure:posterior-crires-tess-cheops}b. Analogous to the high-resolution retrieval in Sect.~\ref{Results from CRIRES$^+$ data and discussion}, the posterior distribution of the temperature parameter $T_1$ flattens towards higher values, resulting in a lower limit on the parameter at $T_1>3200$\,K. This finding is in line with our expectations, since space-based photometry provides little additional information about the physical conditions in the upper atmospheric layers of gas giant exoplanets. 

\begin{figure}
        \centering
        \includegraphics[width=\columnwidth]{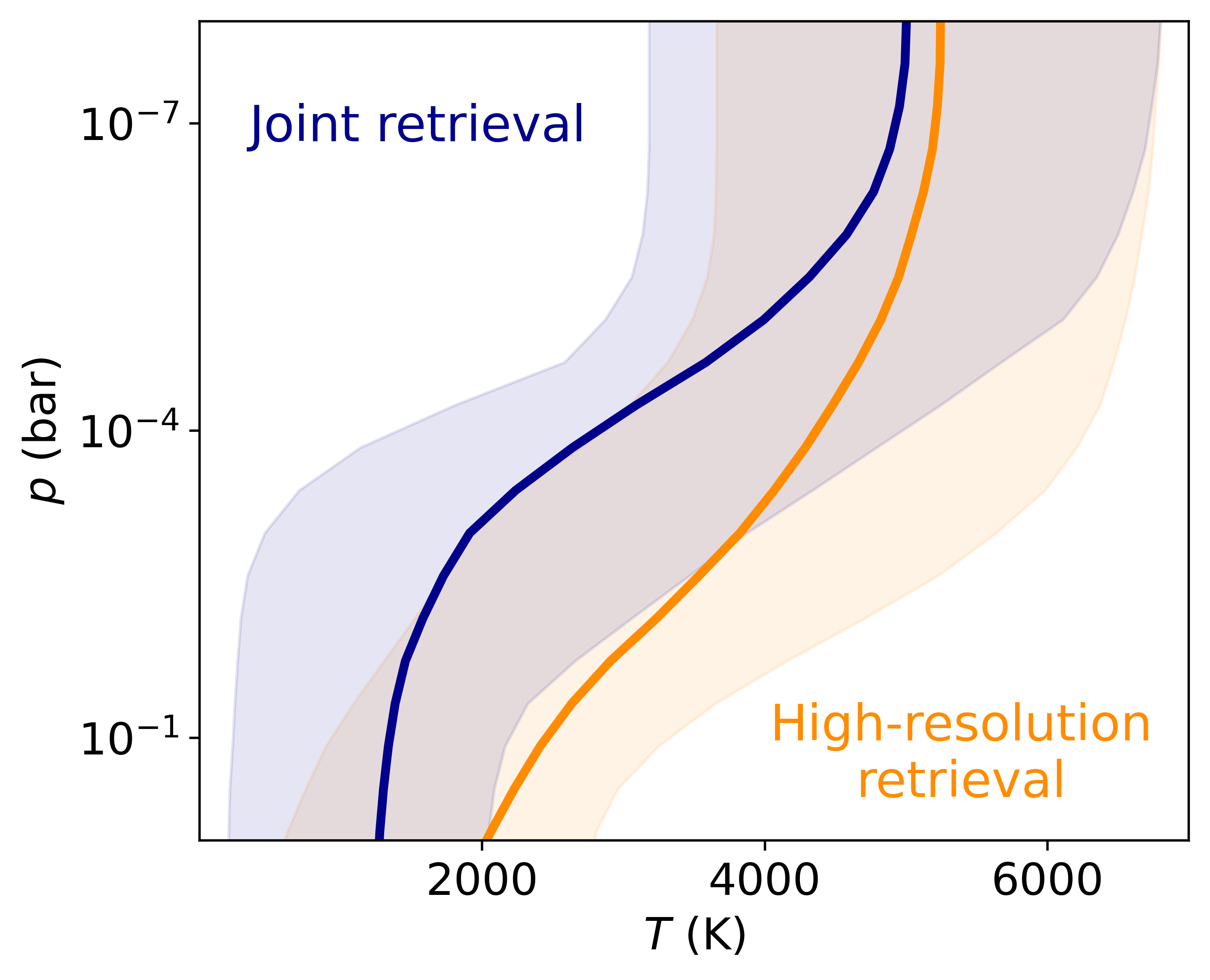}
        \caption{Comparison of retrieved $T$-$p$ profiles. The temperature profile inferred by the high-resolution CRIRES$^+$ retrieval is shown in orange, the temperature profile derived from the joint retrieval with TESS and CHEOPS data is shown in blue. The solid lines correspond to the median, the shaded area to the 95 percentile intervals of the sampled temperature profiles.}
        \label{figure:T-p-comparison}
\end{figure}

The refined constraints on the temperature limit at the bottom of the atmospheric inversion layer of $T_2<2000$\,K are a direct consequence of the inclusion of the TESS and CHEOPS data, which encode the spectral continuum level information from the lower planetary atmosphere. Our improved ability to probe the low-altitude boundary of the thermal inversion is also reflected in the disappearance of the degeneracy between the temperature and pressure parameters $T_2$ and $p_2$. This allows us to measure the thermal properties of the atmosphere more accurately than with high-resolution data alone. Breaking this degeneracy slightly lowers $p_2$, shifting the measured thermal inversion layer to higher atmospheric altitudes. This shift can be seen in Fig.~\ref{figure:T-p-comparison}, which compares the $T$-$p$ profiles derived from the high-resolution retrieval and the joint retrieval with TESS and CHEOPS.

The metallicity of [M/H]\,=\,$1.39_{-1.58}^{+1.64}$\,dex obtained from the joint retrieval is higher than that based on the analysis of the high-resolution observations alone. However, given the relatively large uncertainty intervals, the result is still consistent with the metallicity value of WASP-178b's host star. Our finding of an increased [M/H] value in comparison to the high-resolution retrieval is mainly caused by the thermal inversion layer being shifted to a lower pressure regime. An increase in elemental abundances is required to ensure that the VMRs of $^{12}$CO and H$_2$O remain high enough to maintain the strength of the spectral emission lines at lower pressures. 
Our result is in line with previous reports of atmospheric abundances exceeding solar levels for a number of other UHJs \citep[e.g.,][]{Cont2022b, Kasper2023}. 

The improved characterization of the $T$-$p$ profile through knowledge of the spectral continuum level allows the derivation of bounded constraints on the C/O ratio. 
We determine a value of $0.59_{-0.23}^{+0.20}$, which is consistent with the upper limit obtained in the high-resolution retrieval.  
At visible wavelengths, the continuum level of UHJ spectra is sensitive to the presence of TiO and VO. Consequently, the improved constraints on the C/O ratio are partly due to the contribution of these two molecular species to the emission flux in the TESS and CHEOPS bandpasses. Evidence for depletion of Ti- and V-bearing species, including TiO and VO, from UHJ atmospheres has been found in a number of high-resolution spectroscopy studies \citep[e.g.,][]{Merritt2020, Hoeijmakers2022}. This scenario is not considered in the equilibrium chemistry assumption of our retrieval. Thus, we caution that if TiO and VO are rained out, cold trapped, or otherwise depleted from WASP-178b's atmosphere, our retrieval could yield an overestimated C/O ratio. We tested this scenario by running the retrieval without including the two species, which resulted in a lower C/O ratio of approximately 0.4. Future investigations of WASP-178b's atmospheric chemistry will therefore benefit from high-resolution spectroscopy observations in the visible wavelength range, which will allow to assess the presence and abundances of Ti- and V-bearing chemical species.

The systemic and orbital velocity parameters $K_\mathrm{p}$ and $\varv_\mathrm{sys}$ are in agreement with the values obtained in the high-resolution retrieval. This is due to the fact that the information needed to measure these velocity parameters is primarily encoded in the position of the spectral emission lines, which is not strongly affected by the thermal and chemical properties of the planetary atmosphere. Only a minor offset within the uncertainty intervals of the equatorial rotation velocity $\varv_\mathrm{eq}$ can be observed in comparison to the high-resolution retrieval. This variation can be explained by the inclusion of the photometric data points, shifting the thermal inversion layer from which the emission lines originate towards lower pressures. In this scenario, the spectral lines in the forward model of the high-resolution retrieval originate from regions of higher pressure, while those in the combined retrieval originate from regions of lower pressure. Consequently, the spectral lines of the high-resolution retrieval are affected by stronger pressure broadening than those of the combined retrieval. In order to fit the total broadening of the spectral lines in the observational data, a slightly higher value of $\varv_\mathrm{eq}$ is required for the combined retrieval, although consistent within the uncertainty interval. Overall, we suggest that high-resolution spectroscopy observations alone are sufficient to identify Doppler-shifted spectral signatures and spectral line broadening, and thus to infer the dynamical properties of exoplanet atmospheres.

In addition, the use of photometric data provides us with information on the reflectivity of WASP-178b's dayside atmosphere in the TESS and CHEOPS passbands. Assuming that the reflectivity does not vary significantly with wavelength, we inferred an upper limit on the geometric albedo of $A_\mathrm{g}<0.23$, consistent with previous measurements by \cite{Pagano2023}. The low geometric albedo indicates the absence of reflective clouds in the dayside atmosphere of WASP-178b. This finding is consistent with condensate species being largely absent from the planet's dayside hemisphere due to the elevated atmospheric temperatures. Secondary eclipse measurements of other close-in gas giant exoplanets have shown similar scattering properties to those observed in WASP-178b's atmosphere \citep[e.g.,][]{Mallonn2019, JansenKipping2020, Wong2020, Wong2021, Czesla2023}.

\begin{figure*}
        \centering
        \includegraphics[width=\textwidth]{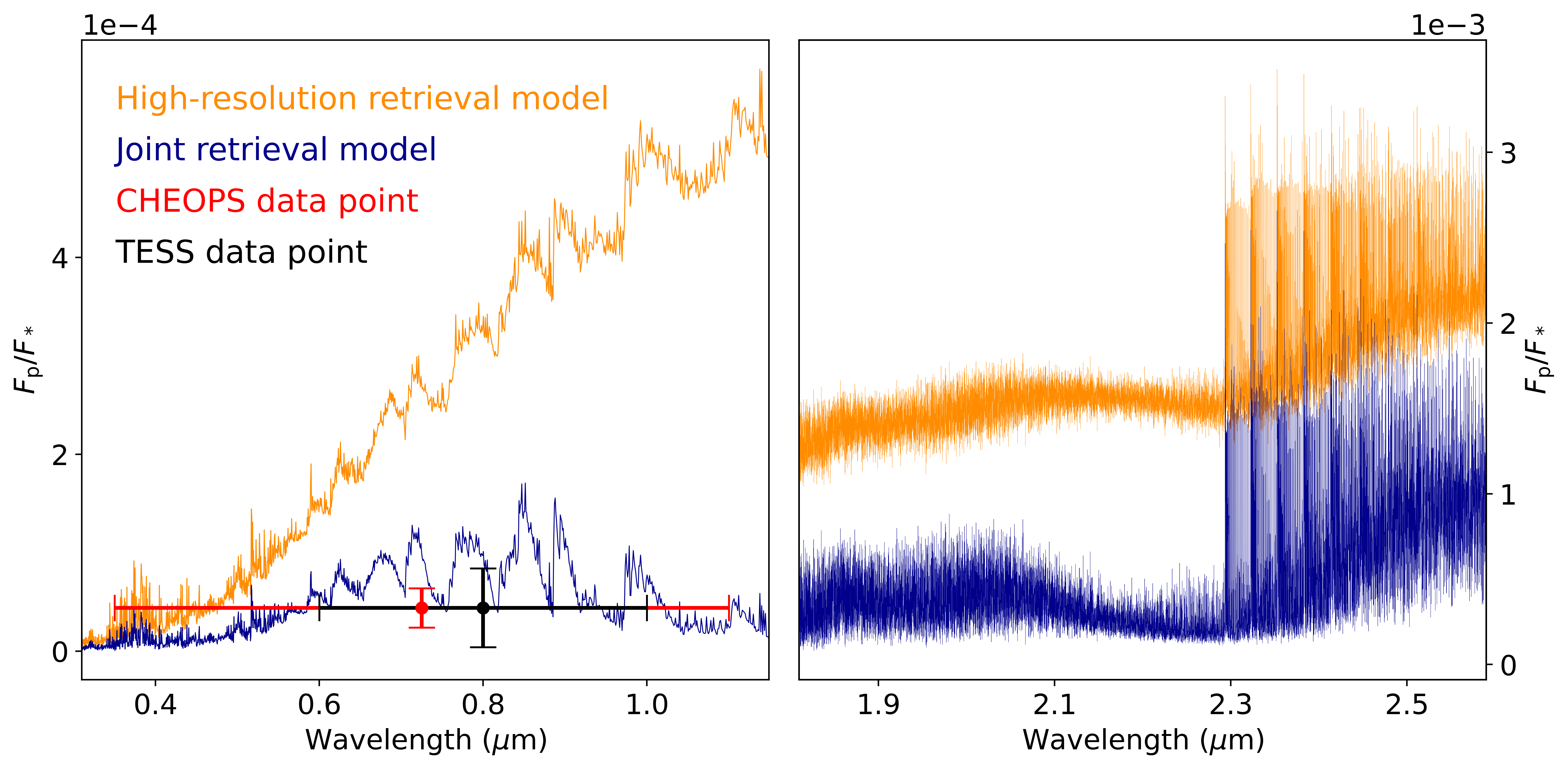}
        \caption{Comparison of model spectra obtained from best-fit parameters of the retrievals. {\it Left panel}: Low-resolution model spectra before integration over the TESS and CHEOPS bandpasses. The parameters derived from the high-resolution retrieval on the CRIRES$^+$ data alone cannot constrain the planetary emission spectrum in the TESS and CHEOPS wavelength range, while the model obtained from the joint retrieval is in agreement with the photometric data points. We note that the plotted eclipse depths are corrected for their reflection component via the retrieved geometric albedo value, since the reported model spectra only consider thermal emission. {\it Right panel}: High-resolution model spectra at CRIRES$^+$ wavelengths. Except for the different continuum level, the two models strongly resemble each other.
        }
        \label{figure:spectra-comparison}
\end{figure*}

Finally, in Fig.~\ref{figure:spectra-comparison} we compare {\tt petitRADTRANS} model spectra calculated with the best-fit parameters of the high-resolution retrieval and the joint retrieval (Table~\ref{tab:retrieval-results}). Apart from a clear difference in the spectral continuum baseline, the two model spectra strongly resemble each other in the CRIRES$^+$ wavelength range, with comparable amplitude and shape of the spectral lines. The high-resolution retrieval can therefore hardly distinguish between the two spectra, since high spectral resolution data only encode the amplitude and shape of the emission lines, but not the continuum level. This explains the limited ability to accurately determine the atmospheric thermal and chemical conditions of WASP-178b with the high-resolution retrieval in Sect.~\ref{Results from CRIRES$^+$ data and discussion}. On the other hand, Fig.~\ref{figure:spectra-comparison} illustrates that incorporating space-based observations from TESS and CHEOPS into the retrieval allows to infer the spectral continuum level, resulting in a best-fit forward model that is in agreement with the photometric data points.

%

\section{Conclusions}
\label{Conclusions}

We have studied the dayside emission spectrum of the UHJ WASP-178b by analyzing high-resolution spectral observations with CRIRES$^+$ in the K band. Using the cross-correlation technique and a Bayesian retrieval framework that includes space-based photometric measurements, the following main results were obtained:

\begin{enumerate}

    \item We identified the spectral signatures of the two molecular species $^{12}$CO and H$_2$O. These are the chemical species with the most dominant emission lines in the spectral K band, and are among the most important carbon- and oxygen-bearing species. No significant signature of the chemical species $^{13}$CO, OH, and \ion{Fe}{i} could be identified in our high-resolution spectroscopy data (Sect.~\ref{Cross-correlation results and discussion}). 
    
    \item The detections of the $^{12}$CO and H$_2$O spectral lines in emission unambiguously prove the existence of a thermal inversion layer in WASP-178b's atmosphere, which is in agreement with theoretical predictions (Sect.~\ref{Cross-correlation results and discussion}). The retrieval on the high-resolution spectroscopy data alone constrains the top and bottom of the thermal inversion layer poorly (Sect.~\ref{Results from CRIRES$^+$ data and discussion}). However, conducting a joint retrieval with secondary eclipse depth measurements from TESS and CHEOPS enables us to obtain tighter uncertainty intervals on the $T$-$p$ profile compared to using CRIRES$^+$ data alone (Sect.~\ref{Joint retrieval with photometric data}). The improved constraints are due to the ability of space-based measurements to capture the information of the spectral continuum level that probes the low-altitude atmospheric layers and the general lack of flux calibration in our high-resolution spectra.

    \item The improved precision on the $T$-$p$ profile obtained from the inclusion of TESS and CHEOPS data points allows a better determination of the atmospheric chemical conditions. The inferred metallicity corresponds to solar to super-solar chemical abundances, which is in line with the metallicity of the host star. The C/O ratio is consistent with the solar value. In particular, a significant improvement in the C/O ratio constraints is achieved by including space-based photometric data in our analysis (Sect.~\ref{Joint retrieval with photometric data}). We emphasize that instead of assuming constant vertical abundances that ignore the presence of thermal dissociation, we approximate the VMRs of the different chemical species by solving for chemical equilibrium in each sampling step of the retrieval (Sect.~\ref{Retrieval framework}). 

    \item The emission lines of $^{12}$CO and H$_2$O show a Doppler-shift relative to each other, indicating the presence of chemical inhomogeneities of the different species in combination with winds across the planetary surface (Sect.~\ref{Cross-correlation results and discussion}). In addition, we infer the presence of strong spectral line broadening that cannot be explained by planetary rotation alone. This excess broadening is most likely caused by a super-rotating atmosphere, the presence of which is generally predicted by general circulation models of UHJs (Sect.~\ref{Results and discussion}).

    \item Finally, the inclusion of photometric data allows us to place an upper limit on the geometric albedo of WASP-178b's dayside atmosphere, in agreement with previous photometric-only measurements (Sect.~\ref{Joint retrieval with photometric data}).

\end{enumerate}

The present study on WASP-178b highlights the potential of combining ground-based high-resolution spectroscopy with spaceborne data to advance our understanding of the physical, chemical, and dynamical properties of UHJ atmospheres. Despite this potential, only a limited number of studies have explored this approach \citep[e.g.,][]{Brogi2017, Gandhi2019, Yan2022a, Boucher2023, Smith2024}. Further advances in atmospheric retrieval techniques will therefore be beneficial to fully exploit the synergies between state-of-the-art ground- and space-based facilities such as CRIRES$^+$ and JWST.

%

\begin{acknowledgements}
    The authors thank the referee for very useful comments and suggestions. 
    CRIRES$^+$ is an ESO upgrade project carried out by Th\"{u}ringer Landessternwarte Tautenburg, Georg-August Universit\"{a}t G\"{o}ttingen, and Uppsala University. The project is funded by the Federal Ministry of Education and Research (Germany) through Grants 05A11MG3, 05A14MG4, 05A17MG2 and the Knut and Alice Wallenberg Foundation. This project is based on observations collected at the European Organisation for Astronomical Research in the Southern Hemisphere under the ESO programme 111.254J.001. 
    D.C. is supported by the LMU-Munich Fraunhofer-Schwarzschild Fellowship and by the Deutsche Forschungsgemeinschaft (DFG, German Research Foundation) under Germany's Excellence Strategy - EXC 2094 - 390783311. 
    F.Y. acknowledges the support by the National Natural Science Foundation of China (grant No. 42375118). 
    F.L. acknowledges the support by the Deutsche Forschungsgemeinschaft (DFG, German Research Foundation) – Project number 314665159. 
    S.C. and M.R acknowledge the support of the DFG priority program SPP 1992 ``Exploring the Diversity of Extrasolar Planets'' (CZ 222/5-1, RE 1664/16-1, and DFG PR 36 24602/41). 
    N.P., L.B.-Ch., and A.D.R. acknowledge support by the Knut and Alice Wallenberg Foundation (grant 2018.0192). 
    E.N. acknowledges the support by the DFG Research Unit FOR2544 ``Blue Planets around Red Stars''. 
    D.S. acknowledges financial support from the project PID2021-126365NB-C21(MCI/AEI/FEDER, UE) and from the Severo Ochoa grant CEX2021-001131-S funded by MCIN/AEI/10.13039/501100011033. 
    This work has made use of the following Python packages: {\tt Astropy} \citep{AstropyCollaboration2013}, {\tt CMasher} \citep{vanderVelden2020}, {\tt corner} \citep{ForemanMackey2016}, {\tt GNU Parallel} \citep{Tange2021}, {\tt Matplotlib} \citep{Hunter2007}, {\tt NumPy} \citep{Harris2020}, {\tt PyAstronomy} \citep{Czesla2019}, and {\tt SciPy} \citep{Virtanen2020}. 
    The authors would also like to thank Paul Molli\`ere for his assistance in selecting the molecular line lists and calculating the high-resolution opacity files for use in {\tt petitRADTRANS}. 
\end{acknowledgements}

%

\bibliographystyle{aa} 
\bibliography{references}

%

\appendix 

\section{Posterior distributions}

\FloatBarrier
\vspace{1.5cm}

\begin{figure*}[h]
        \centering
        \onecolumn        
        \includegraphics[width=\textwidth]{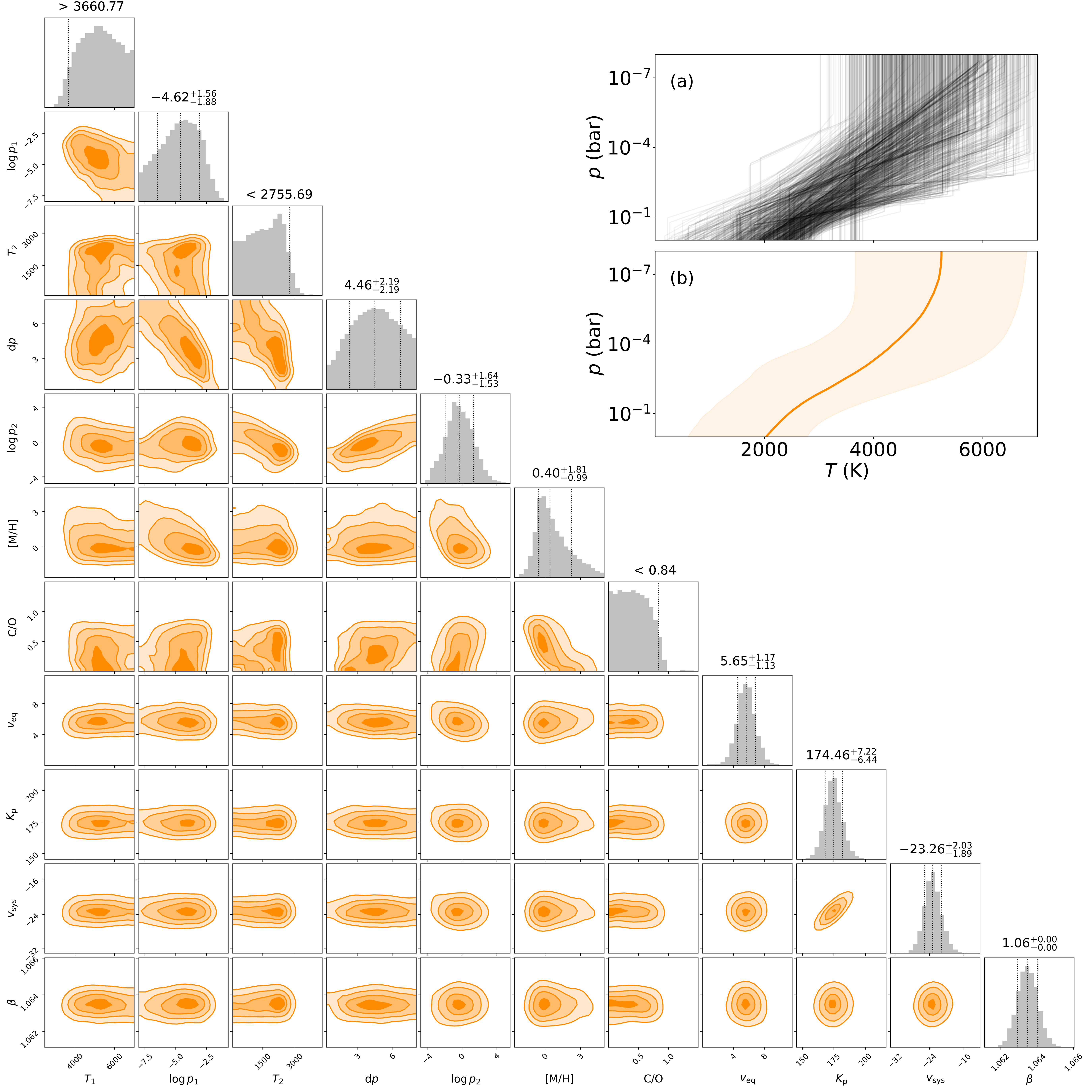}
        \caption{Results of high-resolution retrieval. {\it Corner plot}: posterior distributions and correlations between the atmospheric parameters. The dashed vertical lines in the posterior distributions denote the median and 1$\sigma$ intervals for the bounded parameters. For parameters with upper or lower limits, we report the 2$\sigma$ intervals. {\it Panel a}: examples of the $T$-$p$ profiles sampled by the MCMC analysis, showing that the bottom of the thermal inversion is poorly constrained. {\it Panel b}: median temperature curve with 95 percentiles derived from 3000 random posterior draws.}
        \label{figure:posterior-crires}
\end{figure*}

\clearpage
\vspace*{1.5cm}

\begin{figure*}[h]
        \centering
        \includegraphics[width=\textwidth]{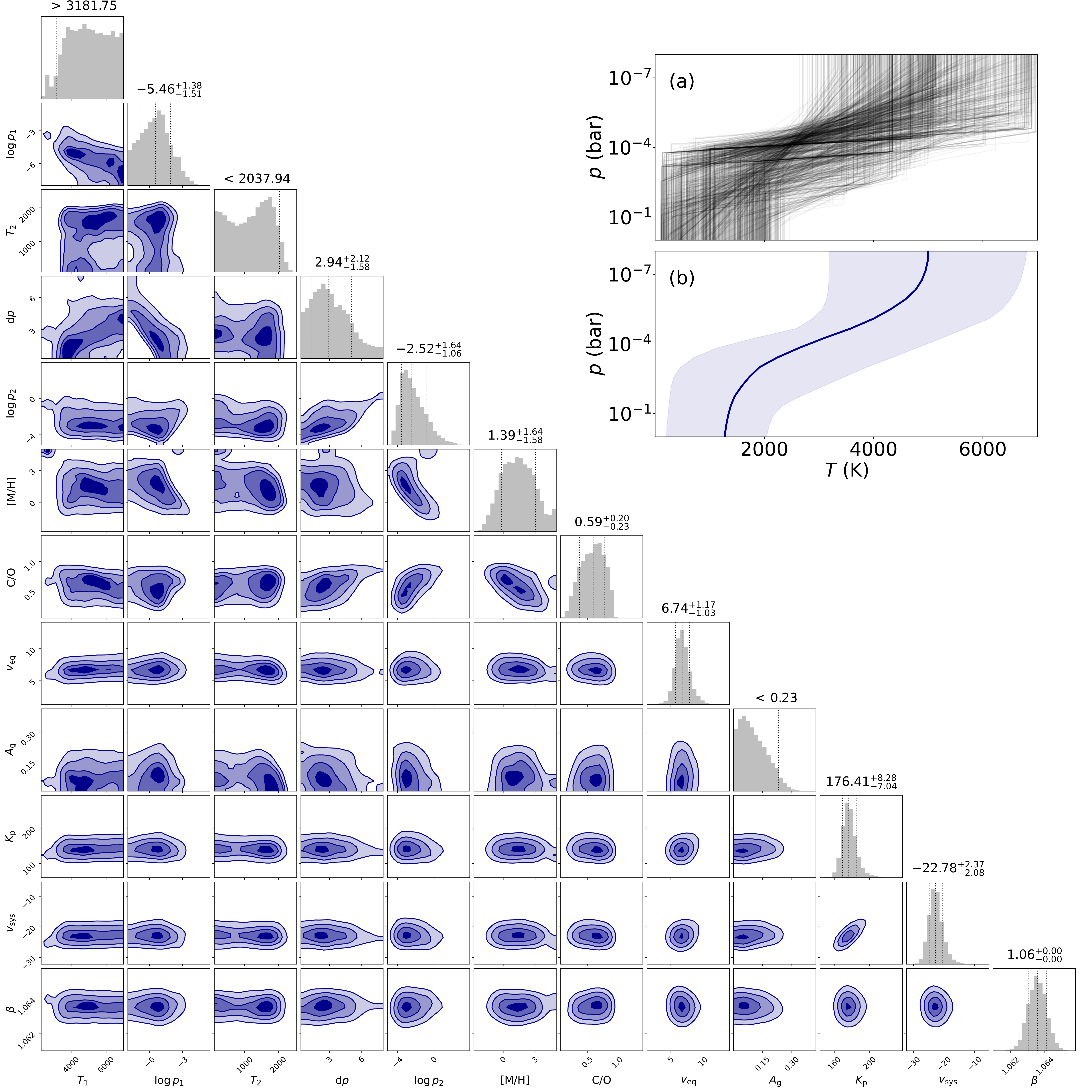}
        \caption{Same as Fig.~\ref{figure:posterior-crires}, but for joint retrieval with photometric data. The $T$-$p$ profiles sampled by the MCMC analysis constrain the thermal inversion better compared to the high-resolution retrieval.}
        \label{figure:posterior-crires-tess-cheops}
\end{figure*}

\end{document}